\PassOptionsToPackage{unicode}{hyperref}
\PassOptionsToPackage{hyphens}{url}
\documentclass[12pt]{article}
\usepackage{amsmath}
\usepackage{graphicx,psfrag,epsf}
\usepackage{enumerate}
\usepackage[]{natbib}
\usepackage{textcomp}

\newcommand{\blind}{0}

\usepackage{color}
\usepackage{fancyvrb}

\DefineVerbatimEnvironment{Highlighting}{Verbatim}{commandchars=\\\{\}}
\usepackage{framed}
\definecolor{shadecolor}{RGB}{248,248,248}
\newenvironment{Shaded}{\begin{snugshade}}{\end{snugshade}}

\newcommand{\AttributeTok}[1]{\textcolor[rgb]{0.13,0.29,0.53}{#1}}

\newcommand{\CommentTok}[1]{\textcolor[rgb]{0.56,0.35,0.01}{\textit{#1}}}

\newcommand{\ConstantTok}[1]{\textcolor[rgb]{0.56,0.35,0.01}{#1}}

\newcommand{\DecValTok}[1]{\textcolor[rgb]{0.00,0.00,0.81}{#1}}

\newcommand{\FunctionTok}[1]{\textcolor[rgb]{0.13,0.29,0.53}{\textbf{#1}}}

\newcommand{\NormalTok}[1]{#1}

\newcommand{\SpecialCharTok}[1]{\textcolor[rgb]{0.81,0.36,0.00}{\textbf{#1}}}

\newcommand{\StringTok}[1]{\textcolor[rgb]{0.31,0.60,0.02}{#1}}

\usepackage{longtable,booktabs,array}
\usepackage{calc} 
\usepackage{etoolbox}
\makeatletter
\patchcmd\longtable{\par}{\if@noskipsec\mbox{}\fi\par}{}{}
\makeatother
\IfFileExists{footnotehyper.sty}{\usepackage{footnotehyper}}{\usepackage{footnote}}
\makesavenoteenv{longtable}

\usepackage{xcolor}
\usepackage{hyperref}
\xdefinecolor{blue}{rgb}{0.28,0.65,0.9}
\definecolor{grey}{rgb}{0.45, 0.45, 0.45}
\hypersetup{
    colorlinks,
    linkcolor={blue},
    citecolor={blue},
    urlcolor={blue}
}
\usepackage{setspace}
\usepackage{multicol}

\usepackage[margin=0.6cm]{caption}
\usepackage{floatrow}
\usepackage{subcaption}
\DeclareNewFloatType{chunk}{placement=H, fileext=chk, name=}
\makeatletter
\@addtoreset{chunk}{section}
\makeatother

\usepackage{amsmath}
\usepackage{booktabs}
\usepackage{longtable}
\usepackage{array}
\usepackage{multirow}
\usepackage{wrapfig}
\usepackage{float}
\usepackage{colortbl}
\usepackage{pdflscape}
\usepackage{tabu}
\usepackage{threeparttable}
\usepackage{threeparttablex}
\usepackage[normalem]{ulem}
\usepackage{makecell}
\usepackage{xcolor}

\IfFileExists{bookmark.sty}{\usepackage{bookmark}}{\usepackage{hyperref}}
\IfFileExists{xurl.sty}{\usepackage{xurl}}{} 
\hypersetup{
  pdftitle={Teaching modeling in introductory statistics: A comparison of formula and tidyverse syntaxes},
  pdfkeywords={R language, instruction, data science, statistical computing},
  hidelinks,
  pdfcreator={LaTeX via pandoc}}

\begin{document}

\def\spacingset#1{\renewcommand{\baselinestretch}%
{#1}\small\normalsize} \spacingset{1}


\if0\blind
{
  \title{\bf Teaching modeling in introductory statistics: A comparison of formula and tidyverse syntaxes}

  \author{
        Amelia McNamara \footnote{\href{mailto:amelia.mcnamara@stthomas.edu}{\nolinkurl{amelia.mcnamara@stthomas.edu}}} \\
    Computer and Data Sciences Department, University of St Thomas\\
      }
  \maketitle
} \fi

\if1\blind
{
  \bigskip
  \bigskip
  \bigskip
  \begin{center}
    {\LARGE\bf Teaching modeling in introductory statistics: A comparison of formula and tidyverse syntaxes}
  \end{center}
  \medskip
} \fi

\bigskip
\begin{abstract}
There are many pedagogical considerations for incorporating programming into a statistics course. When using the programming language R, one consideration is the particular R syntax that will be used. This paper reports on a head-to-head comparison run in a pair of introductory statistics labs, one conducted fully in the formula syntax, the other in tidyverse. Analysis of pre- and post-survey data show minimal differences between the two labs, with students reporting a positive experience regardless of section. Analysis of data from YouTube and RStudio Cloud show interesting distinctions. The formula section appeared to watch a larger proportion of pre-lab YouTube videos, but spend less time computing on RStudio Cloud. Conversely, the tidyverse section watched a smaller proportion of the videos and spent more time computing. Analysis of lab materials showed tidyverse labs tended to be slightly longer in terms of lines in the provided RMarkdown materials and minutes of the associated YouTube videos. The tidyverse labs exposed students to more distinct R functions, but reused functions more frequently. Both labs relied on a relatively small vocabulary of consistent functions, which can provide a starting point for instructors interested in teaching introductory statistics in R. The instructor experience of teaching in the two syntaxes diverged primarily when discussing relationships between categorical variables, as well as when working with summary statistics for numeric variables. This work provides additional evidence for instructors looking to choose between syntaxes for introductory statistics teaching.
\end{abstract}

\noindent%

\vfill

\newpage
\spacingset{1.9} 

\doublespacing

\linespread{1}
\linespread{2}
\vspace{3mm}\setlength{\parindent}{15pt}

\linespread{1}
\linespread{2}
\vspace{3mm}\setlength{\parindent}{15pt}

\section{Introduction}\label{introduction}

When teaching statistics and data science, it is crucial for students to engage authentically with data. The revised Guidelines for Assessment and Instruction in Statistics Education (GAISE) College Report provides recommendations for instruction, including ``Integrate real data with a context and purpose'' and ``Use technology to explore concepts and analyze data'' \citep{carveretal2016}. Many instructors have students engage with data using technology through in-class experiences or separate lab activities.

An important pedagogical decision when choosing to teach data analysis is the choice of tool. There has long been a divide between `tools for learning' and `tools for doing' data analysis \citep{mcnamara2015}. Tools for learning include applets, and standalone software like TinkerPlots, Fathom, or their next-generation counterpart CODAP \citep{konoldmiller2001, finzer2002a, CODAP2021}. Tools for doing are used by professionals, and include software packages like SAS as well as programming languages like Julia, R, and Python.

Many tools for learning were inspired by Rolf Biehler's 1997 paper, ``Software for Learning and for Doing Statistics.'' In it, Biehler called for more attention to the design of tools used for teaching. In particular, he was concerned with on-ramps for students (ensuring the tool was not too complex), as well as off-ramps (using one tool through an entire class, which could also extend further) \citep{biehler1997}. At the time he wrote the paper it was quite difficult to teach using an authentic tool for doing, because these tools lacked technological or pedagogical on-ramps.

However, recent Integrated Development Environments (IDEs) and pedagogical advances have opened space for a movement to teach even novices statistics and data science using programming. In particular, curricula using Python and R have become popular. In these curricula, educators make pedagogical decisions about what code to show students, and how to scaffold it. In both the Python and R communities, there have been movements to simplify syntax for students.

For example, the UC Berkeley Data 8 course uses Python, including elements of the commonly-used \texttt{matplotlib} and \texttt{numpy} libraries as well as a specialized library written to accompany the curriculum called \texttt{datascience} \citep{adhikarietal2021, deneroetal2022}. The \texttt{datascience} library was designed to reduce complexity in the code. At the K-12 level, the language Pyret has been developed as a simplified version of Python to accompany the Bootstrap Data Science curriculum \citep{krishnamurthietal2020}.

In R, the development of less-complex code for students has been under consideration for even longer \citep{pruimetal2011}. R offers non-standard evaluation, which allows package authors to create new `syntax' for their packages \citep{morandatetal2012}. In human language, syntax is the set of rules for how words and sentences should be structured. If someone uses the wrong syntax in human language, people will hear there is something wrong with the speech or writing. However, because human understanding is flexible, the listener will probably still understand the general idea the speaker was trying to convey. Syntax in programming language is more formal-- it governs what code will execute, run, or compile correctly. Using the wrong syntax usually means failing to get a result from the program.

Typically, programming languages have only one valid syntax. For example, an aphorism about Python is ``There should be one-- and preferably only one --obvious way to do it'' \citep{peters2004}. But, non-standard evaluation in R has allowed there to be many obvious ways to do the same task \citep[\citet{wickham2019a}]{morandatetal2012}. There is some disagreement over whether syntax is a precise term for these differences. Other terms suggested for these variations in valid R code are `dialects,' `interfaces,' and `domain specific languages.' Throughout this paper, we use the term syntax as a shorthand for these concepts. At present, there are three primary syntaxes used: base, formula, and \texttt{tidyverse} \citep{mcnamara2018a}.

The base syntax is used by the base R language \citep{rcoreteam2020}, and is characterized by the use of dollar signs and square brackets. The formula syntax uses the tilde to separate response and explanatory variable(s) \citep{pruimetal2017}. The \texttt{tidyverse} syntax uses a data-first approach, and the pipe to move data between steps \citep{wickhametal2019}. In this paper, the base R pipe \texttt{\textbar{}\textgreater{}} has been used, but in the original pedagogical materials students saw the \texttt{magrittr} pipe, \texttt{\%\textgreater{}\%} \citep{bachewickham2014}.

A comparison of using the three syntaxes for univariate statistics and displays can be seen in \ref{r-syntax}. This example code, like the rest in this paper, uses the \texttt{palmerpenguins} data \citep{horstetal2020}. All three pieces of code accomplish the same tasks, and all three use the R language. However, the syntax varies considerably.

\linespread{1}

\begin{Shaded}
\begin{Highlighting}[]
\CommentTok{\# load data}
\FunctionTok{library}\NormalTok{(palmerpenguins)}
\FunctionTok{data}\NormalTok{(}\StringTok{"penguins"}\NormalTok{)}

\CommentTok{\# base syntax}
\FunctionTok{hist}\NormalTok{(penguins}\SpecialCharTok{$}\NormalTok{bill\_length\_mm)}
\FunctionTok{mean}\NormalTok{(penguins}\SpecialCharTok{$}\NormalTok{bill\_length\_mm, }\AttributeTok{na.rm =} \ConstantTok{TRUE}\NormalTok{)}

\CommentTok{\# formula syntax}
\FunctionTok{library}\NormalTok{(mosaic)}
\FunctionTok{gf\_histogram}\NormalTok{(}\SpecialCharTok{\textasciitilde{}}\NormalTok{bill\_length\_mm, }\AttributeTok{data =}\NormalTok{ penguins)}
\FunctionTok{mean}\NormalTok{(}\SpecialCharTok{\textasciitilde{}}\NormalTok{bill\_length\_mm, }\AttributeTok{data =}\NormalTok{ penguins, }\AttributeTok{na.rm =} \ConstantTok{TRUE}\NormalTok{)}

\CommentTok{\# tidyverse syntax}
\FunctionTok{library}\NormalTok{(tidyverse)}
\FunctionTok{ggplot}\NormalTok{(penguins) }\SpecialCharTok{+}
  \FunctionTok{geom\_histogram}\NormalTok{(}\FunctionTok{aes}\NormalTok{(}\AttributeTok{x =}\NormalTok{ bill\_length\_mm))}
\NormalTok{penguins }\SpecialCharTok{|\textgreater{}}
  \FunctionTok{drop\_na}\NormalTok{(bill\_length\_mm) }\SpecialCharTok{|\textgreater{}}
  \FunctionTok{summarize}\NormalTok{(}\FunctionTok{mean}\NormalTok{(bill\_length\_mm))}
\end{Highlighting}
\end{Shaded}

\captionof{chunk}{Making a histogram of bill length from the penguins dataset, then taking the mean, using three different R syntaxes. Base syntax is characterized by the dollar sign, formula by the tilde, and tidyvese is dataframe-first.}

\label{r-syntax}
\linespread{2}
\vspace{3mm}\setlength{\parindent}{15pt}

There is some agreement about pedagogical best practices while teaching R. In particular, most educators agree that in order to reduce cognitive load, instructors should only teach one syntax, and should be as consistent as possible about that syntax \citep{mcnamaraetal2021}. There is also some agreement that base R syntax is not the appropriate choice for introductory statistics, but there is widespread disagreement on whether the formula syntax or \texttt{tidyverse} syntax should be used in these courses.

While there are strongly-held opinions on which syntax should be taught \citep{pruimetal2017, cetinkaya-rundeletal2022}, there is relatively little empirical evidence to support these opinions. In the realm of computer science, empirical studies have shown significant differences in the intuitiveness of languages, as well as error rates, based on language design choices \citep{stefiketal2011, stefiksiebert2013}. Thus, it seems likely there are language choices that could make data science programming easier (or harder) for users, particularly novices.

Rafalski et al ran an experiment comparing the three main R syntaxes. The study showed no statistically significant difference between the three syntaxes with regard to time to completion or number of errors. However, there were significant interaction effects between syntax and task, which suggested some syntaxes might be more appropriate for certain tasks \citep{rafalskietal2019}.

Examining the results of the study with an eye toward data science pedagogy showed common errors related to conceptions of dataframes and variables. For example, one of the figures from \citet{rafalskietal2019} shows real student code with errors. In the first line of code, the student gets everything correct using formula syntax, with the exception of the name of the dataframe. When that code does not work, they try again using base R syntax, but again get the dataframe name wrong. After both those failures, they appear to fall back on computer science knowledge and try syntax quite different from R. This is consistent with other studies of novice behavior in R \citep{roberts2015}. It is not clear if this type of error was dependent on the syntax participants were asked to use.

The study was a quick intervention showing students examples of a particular syntax, then asking them to duplicate that syntax in a new situation. But without any instruction about data science concepts like dataframes, it would be difficult to troubleshoot the syntax error mentioned above. The work served as the inspiration for the longer comparison of multiple R syntaxes in the classroom context described in this paper.

In this paper, we seek to more deeply compare the experience of teaching and learning R in an introductory statistics class using the formula and \texttt{tidyverse} syntaxes. This manuscript does not provide an answer to which syntax is `best,' but instead helps to describe the benefits and drawbacks of each syntax. This paper provides instructors with additional questions to ask themselves and evidence to use in service of making a decision on syntax.

The remainder of this paper is organized into four sections. Section \ref{sec:methods} describes the setup of the classes used in the comparison (\ref{sec:structure}), the participants (\ref{sec:participants}), and the content of the course under investigation (\ref{sec:materials}), including the RMarkdown documents provided to students (\ref{sec:rmd}) and the pre-lab YouTube videos for the flipped class (\ref{sec:videolengths}). Section \ref{sec:results} contains results of the pilot study, including a discussion about the lack of summative assessments (\ref{sec:assessment}), results from the pre- and post-survey (\ref{sec:prepost}), analysis of YouTube (\ref{sec:yt}) and RStudio Cloud (\ref{sec:rstudio}) data, a study of the functions shown in each section (\ref{sec:numfunc}), and a qualitative assessment of divergent labs (\ref{sec:diflabs}). Finally, Section \ref{sec:discussion} discusses the results and opportunities for future study.

\section{Methods}\label{sec:methods}

The author conducted a head-to-head comparison of the formula and \texttt{tidyverse} syntaxes in her introductory statistics labs. The comparison was run twice, once in the Spring 2020 semester and once in the Fall 2020 semester. The disruption of COVID-19 to the Spring 2020 semester made the resulting data unusable, so this paper focuses on just Fall 2020 data.

Data was collected from YouTube analytics for watch times, from RStudio Cloud for compute time, and from pre- and post-surveys of students. Participants for the pre- and post-survey were recruited from this pool after Institutional Research Board ethics review (University of St Thomas IRB 1605810-2).

\subsection{Course structure}\label{sec:structure}

This comparison was run in two introductory statistics labs at a mid-sized private university in the upper Midwest. At this university, statistics students enroll in a lecture (approximately 60-90 students per section), which is broken into several smaller lab sections for hands-on work in statistical software. Lecture and lab sections are taught by different instructors, and the lab sections associated with a particular lecture section often use different software. For example, a lecture section of 90 students may have one lab section of 30 students using Minitab while the other two sections use Excel. However, every lab section (no matter what lecture it is associated with, or what software is used) does the same set of standardized assignments. This structure provides a consistent basis for comparison.

In Fall 2020, the author taught two labs sections associated with the same lecture section, meaning all students saw the same lecture content. (A third lab section was associated with the same lecture, using a different software, and was not considered.) Using random assignment (coin flip), the author selected one lab section to be instructed using formula syntax, and one to be instructed using \texttt{tidyverse} syntax. The goal was to compare syntaxes head-to-head.

The lab took place during the coronavirus pandemic, and the lab instructor used a flipped classroom format. Each week, the instructor prepared a ``pre-lab'' document in RMarkdown. The pre-lab covered the topics necessary to complete the standardized lab assignment done by all students across lab sections. Pre-lab documents included text explanations of statistical and R programming concepts, sample code, and blanks (both in the code and the text) for students to fill in as they worked. Students in both sections used R through the online platform RStudio Cloud \citep{rstudiopbc2021}.

The instructor recorded YouTube videos of herself working through the pre-lab documents for each lab, and posted them in advance. Students were expected to watch the pre-lab video(s) and work through the RMarkdown document on their own time, then come to synchronous class to ask questions and get help starting on the lab assignment.

Literature about flipped classrooms suggests shorter videos are better for student engagement, although there is no consensus about the ideal length for videos, with suggestions ranging from 5 to 20 minutes as a maximum length for a video \citep{zuber2016, beattyetal2019, guoetal2014}. The instructor attempted to keep the total number of minutes of video content below 20 each week. If video content became too long, the instructor split the content into multiple shorter videos.

\subsection{Participants}\label{sec:participants}

Participants were students enrolled in two introductory statistics lab sections associated with a common introductory statistics course. The two labs were of the same size (\(n=21\) in both sections) and reasonably similar in terms of student composition. In both sections, approximately half of students were business majors, with the other half a mix of other majors.

Participants for the pre- and post-survey were recruited from this pool after Institutional Research Board ethics review. For the pre-survey, \(n=12\) and \(n=13\) students consented to participate, and in the post-survey \(n=8\) and \(n=13\) responded. So, for paired analysis we have \(n=8\) for the formula section, and \(n=13\) for the \texttt{tidyverse} section. These sample sizes are very small, and because students could opt-in, may suffer from response bias. However, because we have additional usage data from non-respondents, some elements of the data analysis include the full class sample sizes of \(n=21\).

\linespread{1}
\linespread{2}
\vspace{3mm}\setlength{\parindent}{15pt}

\linespread{1}
\linespread{2}
\vspace{3mm}\setlength{\parindent}{15pt}

This lab course accompanied an introductory statistics course, a first introduction to statistics for all but a handful of students. Typically, students also have no prior experience programming. In the formula section, 10 had no prior experience programming, and 2 had experience programming, but not with R. In the \texttt{tidyverse} section, 9 had no prior experience programming, and 4 had experience programming, but not with R. While two additional students in the \texttt{tidyverse} section had prior programming experience, the majority of students in both sections had no prior programming experience.

\linespread{1}
\linespread{2}
\vspace{3mm}\setlength{\parindent}{15pt}

For the students who had programmed before, none had prior experience with R. Three students had prior experience with Java, three with Javascript, and a smaller number had experience with other languages, including C++ and Python.

\subsection{Topics and materials}\label{sec:materials}

The course these labs accompanied was an introductory statistics course using the Lock5 textbook \citep{locketal2020}. Weekly lab topics were aligned with lecture content, and standardized across all labs for the semester (13 lab sections using a variety of different software). The topics covered are shown in Table \ref{topics}.

\begin{table}
\singlespacing
{\setlength{\extrarowheight}{0.1cm}
\begin{tabular}{p{1cm}|p{15cm}}
Week & Topic \\
1 & \em{No lab, short week} \\  
2 & Describing data: determining the number of observations and variables in a dataset, variable types. \\  
3 & Categorical variables: exploratory data analysis for one or two categorical variables. Frequency tables, relative frequency tables, bar charts, two-way tables, and side-by-side bar charts. \\  
4 & Quantitative variables: exploratory data analysis for one quantitative variable. \newline Histograms, dot plots, density plots, and summary statistics like mean, median, and standard deviation. \\  
5 & Correlation and regression: exploratory data analysis for two quantitative variables. Correlation, scatterplot, simple linear regression as a descriptive technique. \\  
6 & Bootstrap intervals: the use of the bootstrap to construct non-parametric confidence intervals. \\  
7 & Randomization tests: the use of randomization to perform non-parametric \linebreak hypothesis tests. \\  
8 & Inference for a single proportion: use of the normal distribution to construct 
\linebreak confidence intervals and perform hypothesis tests for a single proportion. \\  
9 & Inference for a single mean: use of the t-distribution to construct confidence intervals and perform hypothesis tests for a single mean. \\  
10 & Inference for two samples: use of distributional approximations (normal or t) to perform inference for a difference of proportions or a difference of means.\\  
11 & \em{No lab, assessment}\\  
12 & \em{No lab, Thanksgiving}\\  
13 & ANOVA: inference for more than two means, using the F distribution.\\  
14 & Chi-square: inference for more than two counts, using the $\chi^2$ distribution.\\  
15 & Inference for regression: inference for the slope coefficient in simple linear regression, prediction and confidence intervals. Multiple regression. \\  
Finals & \em{Second assessment} \\  
\end{tabular}
}
\caption{List of topics for labs across the 15-week semester.}\label{topics}
\end{table}
\doublespacing

Although this was a 15-week semester, there are only 12 lab topics. Labs were not held during the first week of classes or during Thanksgiving week. Additionally, there were two ``lab assessments'' to gauge student understanding of concepts within the context of their lab software. One took place during finals week, the other was scheduled in week 11.

\subsubsection{RMarkdown documents}\label{sec:rmd}

\begin{figure}

\subfloat[Length of pre-lab RMarkdown documents each week, in lines. Data has been adjusted for the formula section in weeks 8 and 9, because an instructor error led this section to have only one document combining both weeks' work.\label{fig:prelablength-1}]{\includegraphics[width=0.8\linewidth]{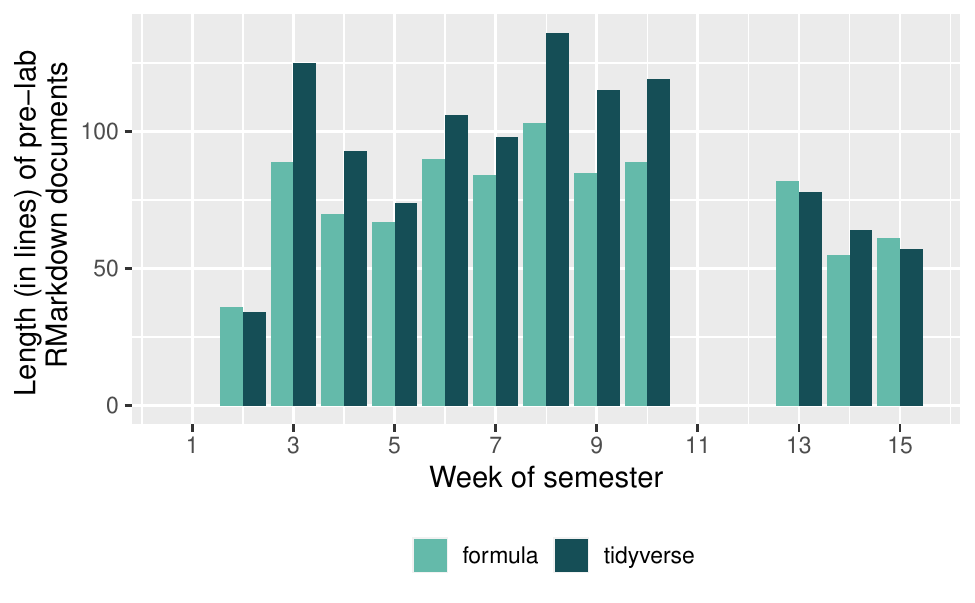} }\newline\subfloat[Length of pre-lab videos each week. Outlines delineate multiple videos for a single week.\label{fig:prelablength-2}]{\includegraphics[width=0.8\linewidth]{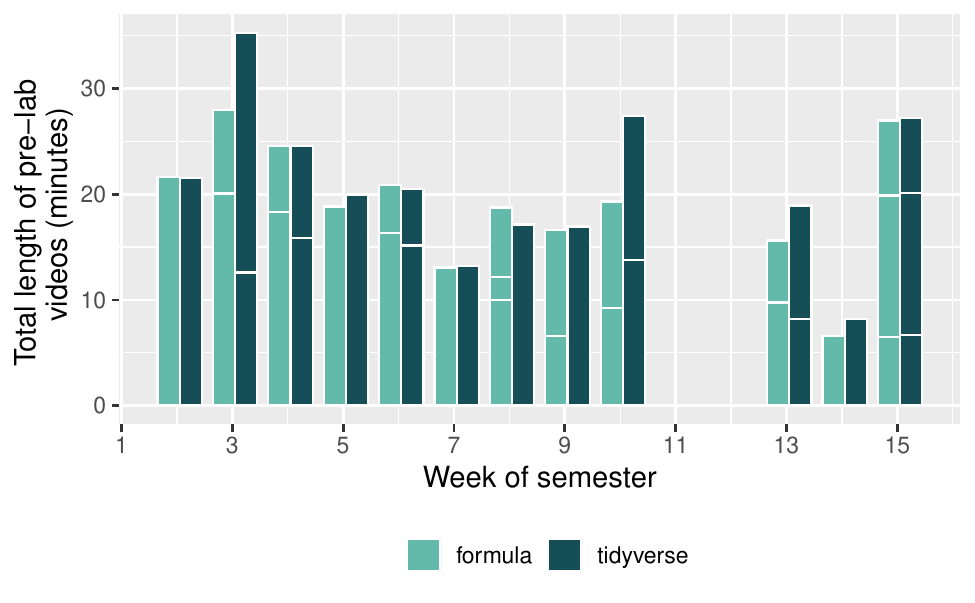} }\newline\hfill{}

\caption{Length of pre-lab documents, in lines and minutes.}\label{fig:prelablength}
\end{figure}

The instructor prepared weekly RMarkdown documents for students, as described in Section \ref{sec:structure}. One question is whether the materials presented to students were of approximately the same length. We can first assess this using the length of the pre-lab RMarkdown documents, measured using lines. Figure \ref{fig:prelablength}a shows the number of lines of code and text in each section's pre-lab document, per week. Lines in the RMarkdown document include the YAML header (consistent between documents), the descriptive text about processes (largely similar between documents), and the code in code chunks, which varied based on the syntax of the lab.

\linespread{1}
\linespread{2}
\vspace{3mm}\setlength{\parindent}{15pt}

Every attempt was made to align these RMarkdown documents, so the descriptive text was only changed when necessary to describe specific elements of the code. Similarly, if blank code chunks appeared in one lab, that was mirrored by a blank chunk in the other lab. Both labs' documents were styled using the \texttt{styler} package \citep{mullerwalthert2022} to remove inconsistencies with spacing, assignment operators, and the like. The \texttt{styler} package is based on the \texttt{tidyverse} style guide \citep{wickham2022}, and has become the de facto style guide for R. Previously-existing style guides like the Google R Style Guide have largely decided to follow \citet{wickham2022}.

Figure \ref{fig:prelablength}a indicates RMarkdown documents for the \texttt{tidyverse} section tended to be longer. A slightly longer length for \texttt{tidyverse} materials makes sense, because \texttt{tidyverse} code is characterized by multiple short lines strung together into a pipeline with \texttt{\textbar{}\textgreater{}}, while the formula syntax typically has single function calls, sometimes with more arguments. (When students engaged with this material, they were using the \texttt{magrittr} pipe, \texttt{\%\textgreater{}\%}, but code shown here has been adjusted to the new native base R pipe, \texttt{\textbar{}\textgreater{}}. This does not have an impact on the number of lines of code.) The code shown in this paper is styled the same way as the labs, so any code comparison shown (for example, in \ref{r-syntax} or \ref{tally-ex1} and \ref{tidy-tally1}) will show the difference in lines of code between the two syntaxes.

\linespread{1}
\linespread{2}
\vspace{3mm}\setlength{\parindent}{15pt}

We can quantify how much longer the \texttt{tidyverse} documents were, either in terms of absolute lines (\texttt{tidyverse} labs were on average 16 lines longer) or in terms of percent difference (\texttt{tidyverse} labs were, on average, 18\% longer).

\subsubsection{YouTube vidoes}\label{sec:videolengths}

Another major instructional source were the YouTube videos the instructor recorded of herself working through the pre-lab documents for each lab.

Figure \ref{fig:prelablength}b shows the lengths of videos over the course of the semester. Video length appears generally consistent between sections. Effort was made to ensure the maximum video length was approximately 20 minutes, although some weeks had multiple videos.

\linespread{1}
\linespread{2}
\vspace{3mm}\setlength{\parindent}{15pt}

We can again compute the percentage difference in total video length (adding together multiple videos in weeks that had them), and compute the mean of that difference. This shows us that \texttt{tidyverse} labs were 2 minutes longer or 9\% longer than formula videos, on average. The \texttt{tidyverse} labs appear to be slightly longer in terms of lines of code and video length, although these differences are slight.

\section{Results}\label{sec:results}

The results from this work include both quantitative and qualitative evidence. Quantitative results can be derived from the pre- and post-survey in the class (intentionally-collected data), as well as YouTube Analytics and RStudio Cloud usage data (incidental data). We can also quantify the number of functions used in both sections by analyzing the RMarkdown documents as data in their own right. All of these quantitative results show slight differences between sections, which hint toward some of the qualitative differences in experience. We also discuss these qualitative differences by highlighting three topics where the experience teaching and learning the syntaxes diverged.

\subsection{Summative assessments}\label{sec:assessment}

One obvious question arising when considering the comparison of the two syntaxes is whether students performed better in one section or another. The IRB did not cover examining student work (an obvious place for improved further research), so we cannot look at student outcomes on a per-assignment basis. However, running a randomization test for a difference in overall mean lab grades showed no significant difference between the two sections. These grades were raw scores, and were not curved in any way. Lab grades comprise 30\% of overall lecture course grade. While there may have been interesting differences in grades depending on the topic of the lab, we at least know these differences averaged out in the end.

Similarly, it would be interesting to know if student attitudes about the instructor were different from the summative student evaluations completed by all students at the end of the semester. These evaluations are anonymous, and the interface only provides summary statistics. Again, a randomization test for a difference in means showed no difference in mean evaluation score on the questions ``Overall, I rate this instructor an excellent teacher.'' and ``Overall, I rate this course as excellent.''

While we cannot make conclusions about summative assessments, there are a number of other data sources to give insight into the difference between the experience of teaching and learning the two syntaxes.

\subsection{Pre- and post-survey}\label{sec:prepost}

Students in both sections were given a pre- and post-survey on their experience in the class. The majority of the survey was modeled on a pre- and post-survey used by The Carpentries, a global nonprofit teaching coding skills \citep{thecarpentries2021}.

As discussed in Section \ref{sec:participants}, the number of students who completed both the pre- and post-surveys was low (\(n=8\) for the formula section, and \(n=13\) for the \texttt{tidyverse} section), so there are major limitations to paired analysis.

On the survey, respondents were asked to use a 5-step Likert scale, from 1 (strongly disagree) to 5 (strongly agree) to rate their agreement with the following statements:

\begin{itemize}
\itemsep-3mm
\linespread{1}
\item I am confident in my ability to make use of programming software to work with data 
\item Having access to the original, raw data is important to be able to repeat an analysis 
\item Using a programming language (like R) can make me more efficient at working with data 
\item While working on a programming project, if I get stuck, I can find ways of overcoming the problem 
\item Using a programming language (like R) can make my analysis easier to reproduce 
\item I know how to search for answers to my technical questions online \linespread{2}
\vspace{3mm}\setlength{\parindent}{15pt}
\end{itemize}

\begin{figure}

{\centering \includegraphics[width=0.95\linewidth]{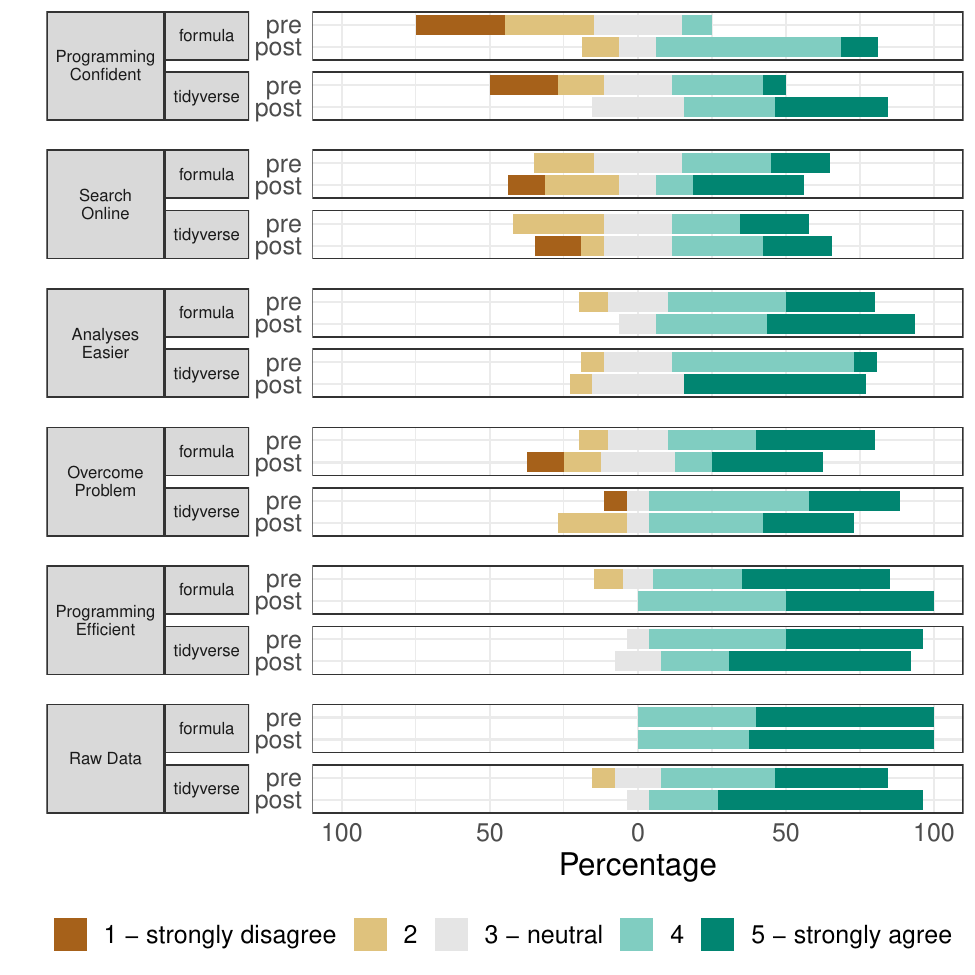} 

}

\caption{Pre- and post-survey responses to Likert-scale questions. Most questions show some level of improvement, such as the first question, `I am confident in my ability to make use of programming software to work with data.' but others show no change or even a decline in agreement.}\label{fig:pre-post}
\end{figure}

Figure \ref{fig:pre-post} displays a visualization of these Likert-scale questions.The visualization shows the pre- and post-survey responses to the survey questions, broken down by section. There is no obvious overall trend. However, considerable improvement was seen in the Programming Confident category, which suggests that students in both sections became more confident in their ``ability to make use of programming software to work with data.''

Other questions showed small gains or unclear changes, like the Analyses Easier category. In the formula section, this question appears to show a small increase in agreement from pre- to post-survey, while the \texttt{tidyverse} section shows either no change or perhaps a slight improvement (more students saying they strongly agree, but fewer in the positive answers overall). Other questions show a decrease in agreement from the pre- to the post-survey, like Overcome Problem. Finally, some questions show improvement in one section but a decline in agreement in the other, like Search Online. In the \texttt{tidyverse} section, there appears to be an increase in their confidence in searching ``for answers to {[}their{]} technical questions online,'' whereas in the formula section there is a decrease.

\linespread{1}
\linespread{2}
\vspace{3mm}\setlength{\parindent}{15pt}

\linespread{1}
\linespread{2}
\vspace{3mm}\setlength{\parindent}{15pt}

Likely, the questions used by The Carpentries were less appropriate for this setting, and a different set of survey questions would have been more appropriate for this group. For example, this class did not include any explicit instruction on searching for answers online. This was an intentional choice, because novices typically struggle to identify which search results are relevant to their queries and get overwhelmed by the multitude of syntactic options they encounter. Instead, students with questions were referred to an ``all the R you need'' cheatsheet they were given at the beginning of the semester, which attempted to summarize every function they would encounter. Likely, students still attempted to search for answers online, which may be why the responses to questions about searching online and overcoming problems got more negative over the course of the semester. In particular, students in the formula section who searched for answers online likely saw unfamiliar syntax, while students in the \texttt{tidyverse} section may have found \texttt{tidyverse} answers because of the popularity of the syntax online.

It is possible that the rise of generative AI systems like ChatGPT will make this easier for students, because they could ask the AI agent to return results only in a specific syntactic style. However, anecdotal experience suggests generative AI systems currently struggle to match less-than-common code styles, like the formula syntax. This is a result of the training data, where \texttt{tidyverse} proliferates and often uses very similar code style, but other styles are more likely to be mixed and matched.

Figure \ref{fig:pre-post} does not utilize the potential for matching pre- and post-responses from the same student to measure change at the individual level. To consider this individual-level change, we can compute the difference between an individual student's response on the pre- and post-survey. We compute \(\text{post score} - \text{pre score}\) such that positive differences mean the student's attitude on the item improved from the beginning of the class to the end, and negative differences mean they worsened.

\linespread{1}
\linespread{2}
\vspace{3mm}\setlength{\parindent}{15pt}

\begin{figure}

{\centering \includegraphics[width=0.8\linewidth]{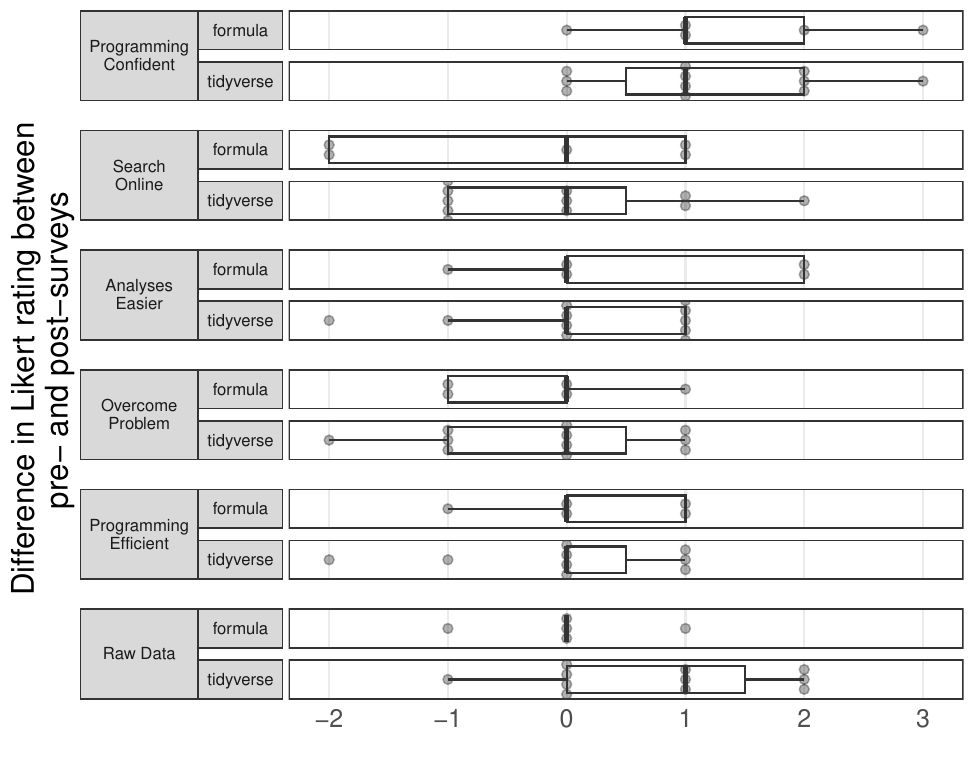} 

}

\caption{Distribution of paired differences for student responses to questions, broken down by section.}\label{fig:prepost}
\end{figure}

Because the questions were on Likert scales, it is not appropriate to compute an arithmetic mean of the differences, but median scores can be computed. To provide a broader picture of the distribution of responses, we also compute the 25th and 75th percentiles for question. This information is most easily displayed as a boxplot. The boxplots in question can be seen in Figure \ref{fig:prepost}.

\linespread{1}
\linespread{2}
\vspace{3mm}\setlength{\parindent}{15pt}

In Figure \ref{fig:prepost}, many of the boxplots are centered at zero (meaning the median response did not change over the course of the semester), so there was no overall difference in medians for those questions.

The one question where the boxes from both sections are centered above 0 is ``I am confident in my ability to make use of programming software to work with data.'' For this question, the median is 1 in both sections, meaning the median student answered one level up on the question at the end of the course. Both boxes (the middle 50\% of the data) are fully positive, although the lower whisker (minimum value) for both still includes zero.

Overall, the boxes in the \texttt{tidyverse} section appear shifted further to the right (more positive) than the formula section, although most of the boxes are still centered at 0. The other question with a positive median for the \texttt{tidyverse} section was ``Having access to the original, raw data is important to be able to repeat an analysis.''

Overall, it seems students improved their confidence in programming over the course of the semester, and there is slight evidence that students in the \texttt{tidyverse} section saw larger improvement in several questions. However, these differences are small and should not be considered statistically significant.

\subsubsection{Additional survey questions}\label{additional-survey-questions}

In addition to the six questions asked on both the pre- and post-survey, the two surveys each had some unique questions.

The pre-survey asked students to share what they were most looking forward to, and most nervous about. Both sections had similar responses. Students wrote they looked forward to ``learning how to code!'' and ``Gaining a better understanding of how to analyze data.'' Beyond worries related to the pandemic, they expressed apprehension about ``getting stuck,'' ``using R,'' and ``Figuring out how to do the programming and typing everything out.''

On the post survey, students were asked to report which syntax they had learned, with an option to respond ``I don't know.'' All students in both sections correctly identified the syntax associated with their lab. Then, they were asked if they would have preferred to learn the other syntax. We hypothesized many students would say `yes,' thinking the other syntax would have been easier or lack some feature they found frustrating. Surprisingly, though, the majority of students in both sections said `no,' they preferred to learn the syntax they had been shown.

\linespread{1}
\linespread{2}
\vspace{3mm}\setlength{\parindent}{15pt}

However, part of the explanation is likely that the students did not know what the other syntax looked like. Throughout the semester, the instructor was careful to only expose students to the syntax for the particular section. Several students asked to see the alternate syntax during office hours, but this was the exception and not the norm.

\linespread{1}
\linespread{2}
\vspace{3mm}\setlength{\parindent}{15pt}

Another question on the post-survey asked students ``How was the experience of learning to program in R?'' Overall, students seem to have positive sentiment toward learning R, whether in the formula or the \texttt{tidyverse} section. As seen in Figure \ref{tab:post-sentiment}, most students said either the experience was ``not what I expected -- in a good way'' or ``About what I expected -- in a good way.''

\linespread{1}
\begin{table}

\caption{\label{tab:post-sentiment}Responses to the question, ``How was the experience of learning to program in R?''}
\centering
\begin{tabular}[t]{llrl}
\toprule
Section & Response & n & Percent\\
\midrule
 & About what I expected -- in a good way & 5 & 24\%\\

 & Not what I expected -- in a good way & 2 & 10\%\\

 & About what I expected -- in a bad way & 1 & 5\%\\

\multirow[t]{-4}{*}{\raggedright\arraybackslash formula} & Not what I expected -- in a bad way &  & \\
\cmidrule{1-4}
 & About what I expected -- in a good way & 3 & 14\%\\

 & Not what I expected -- in a good way & 8 & 38\%\\

 & About what I expected -- in a bad way & 1 & 5\%\\

\multirow[t]{-4}{*}{\raggedright\arraybackslash tidyverse} & Not what I expected -- in a bad way & 1 & 5\%\\
\bottomrule
\end{tabular}
\end{table}\linespread{2}
\vspace{3mm}\setlength{\parindent}{15pt}

Nothing from the survey responses indicated a major difference between the two sections.

\subsection{YouTube analytics}\label{sec:yt}

Because of the flipped format of the class, we can study overall patterns of YouTube watch time. YouTube offers a data portal which allows for date targeting. We defined each week of the semester as running from Sunday to Saturday, covering the time when videos were released (varied dates, typically Monday and Wednesday, since the labs took place Tuesday and Thursday) through to the time finished labs needed to be submitted (Fridays at 11:59 pm). For each week, we downloaded YouTube analytics data for the channel, and filtered the data to focus only on videos related to the introductory statistics labs.

YouTube analytics data includes number of views for each video, number of unique viewers, and total watch time. A ``view'' is defined as a person playing 30 seconds or more of the video, and unique viewers are counted using browser cookies. Data from YouTube is aggregated, and since videos were posted publicly, could contain viewers who were not enrolled in the class. As a way to check for possible inflated view counts from people not in the class, we checked view counts of lab videos on subsequent weeks. For example, we looked at number of views on the the ``describing data'' lab (assigned in week 2) during weeks 3-15. Students in the class would be unlikely to watch videos after the lab's due date, but the general population on the internet would be less targeted in their timing. Rarely did a video garner more than two views in a week that was not the assigned lab week. This indicates there may have been a very small number of non-student views on videos, but they are negligible. We can be reasonably sure the majority of viewers were students.

By limiting the data to a particular week, we were able to join it with data containing the length of the relevant videos. This allows us to calculate the approximate proportion of the videos watched by each student. The data displays some interesting trends.

\begin{figure}

{\centering \includegraphics[width=0.8\linewidth]{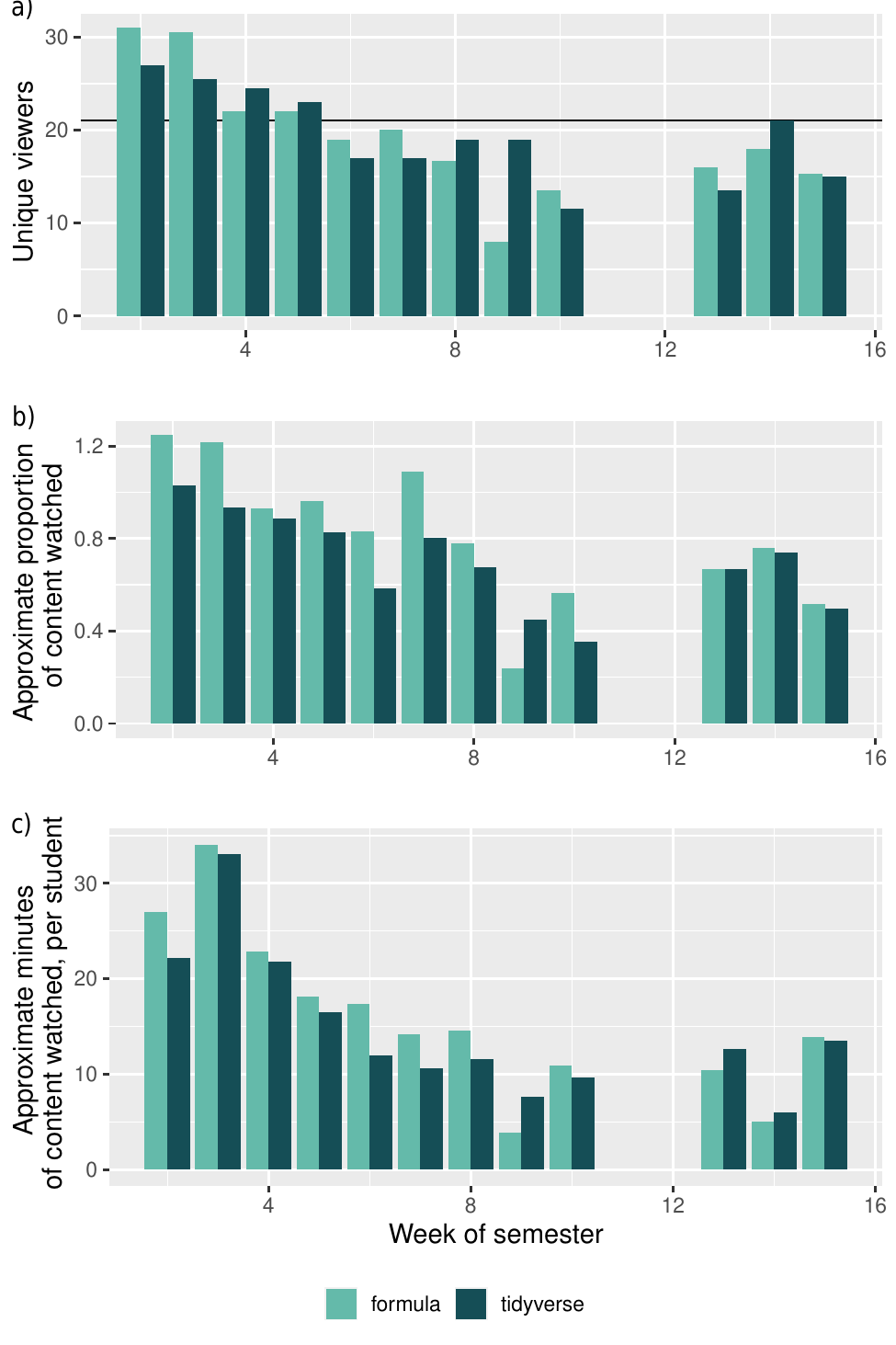} 

}

\caption{Viewing statistics from YouTube. a) shows average number of unique viewers per video. The horizontal line represents the 21 students enrolled in each of the sections, as a baseline for comparison. b) is estimated proportion of YouTube video content watched, per student. This data came from dividing the total amount of time watched by the number of students in each section and the total length of the video(s) for the section that week. c) is estimated number of minutes of YouTube video content watched, per student. This data came from dividing the total amount of time watched by the number of students in each section.}\label{fig:youtube-stats}
\end{figure}

First, we can look at the number of unique watchers per video, seen in Figure \ref{fig:youtube-stats}a. Interestingly, at the start of the semester there are more unique viewers than enrolled students in the class, but as time goes on, the number of unique viewers levels out at slightly less than the number of enrolled students (\(n=21\) for both sections). The lower numbers later on make sense because some students were likely unengaged, or found it possible to do their lab work without watching the videos. However, the high numbers at the start of the semester are puzzling. Perhaps students were viewing the videos from a variety of devices (phone, laptop, computer at school, etc) and therefore being counted as multiple viewers because of different devices or cookie settings such as adblockers or private browsing.

\linespread{1}
\linespread{2}
\vspace{3mm}\setlength{\parindent}{15pt}

\linespread{1}
\linespread{2}
\vspace{3mm}\setlength{\parindent}{15pt}

\linespread{1}
\linespread{2}
\vspace{3mm}\setlength{\parindent}{15pt}

If we assume all viewers were actually students, we can find an approximate proportion of video content watched per student. This is shown in Figure \ref{fig:youtube-stats}b. It appears the proportion of video content watched is larger for the formula videos than for the \texttt{tidyverse} videos. We can characterize the difference by doing pairwise differences of proportion of video watched for each week. The mean of this difference is 11, indicating that on average the formula section watched approximately 11 percentage points more of the videos each week.

One possible reason for this discrepancy is the \texttt{tidyverse} videos tended to be slightly longer, as seen in Section \ref{sec:videolengths}. To explore this, we can also examine the approximate number of minutes of video content watched, per student. This is shown in Figure \ref{fig:youtube-stats}c.~Even though the videos for their section were slightly shorter, it appears students in the formula section watched more minutes of the videos, as well, with a few exceptions.

No matter the explanation, this trend is particularly interesting when considered in conjunction with the RStudio Cloud usage patterns in the following section.

\subsection{RStudio Cloud usage}\label{sec:rstudio}

\linespread{1}
\linespread{2}
\vspace{3mm}\setlength{\parindent}{15pt}

\linespread{1}
\linespread{2}
\vspace{3mm}\setlength{\parindent}{15pt}

Another source of unexpected data was RStudio Cloud usage logs. It was possible to use browser developer tools to download per-student usage data. This data includes all students (\(n=42\)) enrolled in the two sections.

\begin{figure}

{\centering \includegraphics[width=0.8\linewidth]{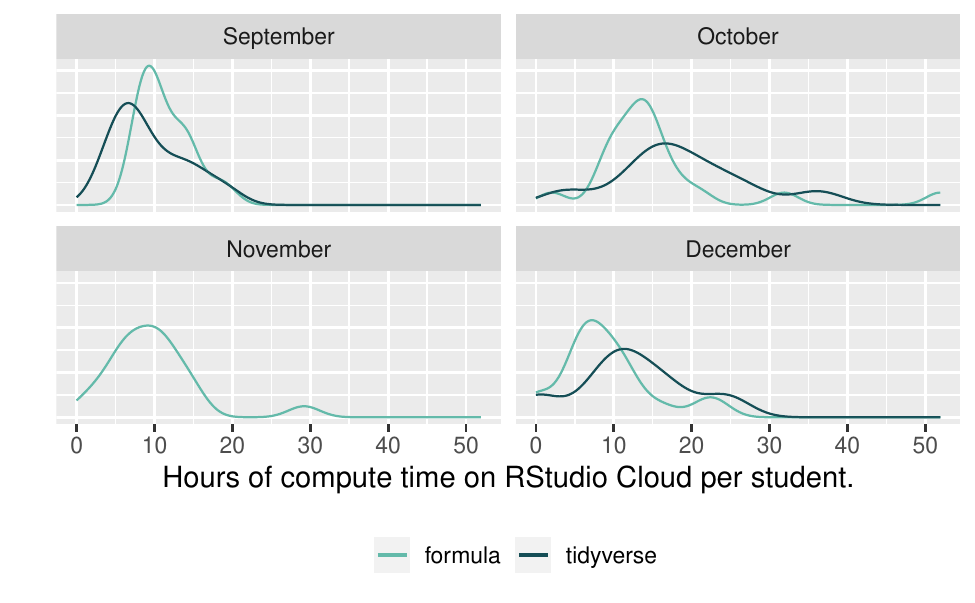} 

}

\caption{Hours of compute time each section spent on RStudio Cloud, per month of the semester. Students in the \texttt{tidyverse} section appear to be spending more time on RStudio Cloud, particularly in the months of October and December.}\label{fig:rstudio-month}
\end{figure}

This allowed us to create Figure \ref{fig:rstudio-month}, which shows the distribution of compute time per student in each section, broken down by month. All the distributions in Figure \ref{fig:rstudio-month} are right-skewed, with several students using many more hours of compute time than the majority. While the \texttt{tidyverse} section seemed to watch less of the provided videos each week (as discussed in Section \ref{sec:yt}), they appeared to spend more time on RStudio Cloud per month.

\linespread{1}
\linespread{2}
\vspace{3mm}\setlength{\parindent}{15pt}

It is important to note these numbers are likely inflated based on the way RStudio Cloud counts usage time. Student projects in both sections were allocated 1 GB of RAM and 1 CPU, so one hour of clock time counted as one project hour (spaces with more RAM or CPU may consume more than one project hour per clock hour), but student usage often includes a fair amount of idle time. RStudio Cloud will put a project to sleep after 15 minutes without interaction, and based on observation of student habits it is likely almost every session ends with a 15 minute idle time before the project sleeps. In a month with four labs, this could add up to an hour of time that does not correspond to students actually using R.

\linespread{1}
\linespread{2}
\vspace{3mm}\setlength{\parindent}{15pt}

Nevertheless, because the numbers would be inflated in the same way in both sections, we will proceed in comparing them. Using data from the entire semester, students in the \texttt{tidyverse} section had a mean number of compute hours per month of 13.5 and students in the formula section had a mean of 11.5 hours. However, there was variability in amount of compute time used per month.

Average use per student per month can be seen in Table \ref{tab:approximate-compute}. The mean compute time for both sections increases from September to October. This makes sense, because only two labs were due in September (an introductory lab with minimal coding, and the first lab on categorical variables), whereas five labs were due in October (weeks 4-8 in Table \ref{topics}). Compute time then drops down in November for the formula section, which makes sense because only two labs and an assessment were due in November (weeks 9-12 in Table \ref{topics}). November data is missing for the \texttt{tidyverse} section. Compute time in December is lower still for the formula section, and also appears lower for the \texttt{tidyverse} section. There were three labs due in December (weeks 13-15 in Table \ref{topics}), as well as an assessment and a final write-up.

Although the labs and assessments were of varying difficulty, we can compute a very crude ``time per assignment'' by dividing the average number of compute hours per month by the number of assignments due in that month, as seen in the final column of Table \ref{tab:approximate-compute}.

\linespread{1}
\linespread{2}
\vspace{3mm}\setlength{\parindent}{15pt}

\linespread{1}
\begin{table}

\caption{\label{tab:approximate-compute}RStudio Cloud usage statistics. Average use per student per month, broken down by section. To find approximate time per assignment, we have divided each month's average by the number of assignments due in the month. The difference between the sections is also computed, and converted into minutes.}
\centering
\begin{tabular}[t]{ll>{\raggedright\arraybackslash}p{1.2in}>{\raggedright\arraybackslash}p{0.9in}>{\raggedright\arraybackslash}p{1.5in}}
\toprule
 &  & mean compute time (sd) & number of assignments & approximate time per assignment\\
\midrule
 & formula & 11.4 (3.3) & 2 & 5.69\\

 & tidyverse & 9.4 (4.7) & 2 & 4.7\\

\multirow[t]{-3}{*}{\raggedright\arraybackslash September} & difference &  &  & -0.99 (-59 minutes)\\
\cmidrule{1-5}
 & formula & 15.7 (10.3) & 5 & 3.15\\

 & tidyverse & 18.7 (8.6) & 5 & 3.73\\

\multirow[t]{-3}{*}{\raggedright\arraybackslash October} & difference &  &  & 0.58 (35 minutes)\\
\cmidrule{1-5}
 & formula & 9.7 (6) & 3 & 3.22\\

 & tidyverse & missing & 3 & missing\\

\multirow[t]{-3}{*}{\raggedright\arraybackslash November} & difference &  &  & missing\\
\cmidrule{1-5}
 & formula & 9.1 (6) & 5 & 1.82\\

 & tidyverse & 12.3 (7.2) & 5 & 2.46\\

\multirow[t]{-3}{*}{\raggedright\arraybackslash December} & difference &  &  & 0.64 (38 minutes)\\
\bottomrule
\end{tabular}
\end{table}\linespread{2}
\vspace{3mm}\setlength{\parindent}{15pt}

While students in the formula section spent more time computing in September, in the following months it appears students in the \texttt{tidyverse} section spent more time on RStudio Cloud. We can concoct several scenarios to explain this difference. In one, students in the \texttt{tidyverse} section were more engaged with their work, so spent more time playing with code in R. In another, students in the \texttt{tidyverse} section struggled to complete their work, so spent more time in R trying to get their lab material to work. A more neutral third option is just that some of the tasks take more code to accomplish (as discussed in \ref{sec:diflabs}), so students needed more time to do their work. Because the usage data was collected incidentally after the fact, we have no information about which story is closer to the truth. A follow-up study might conduct semi-structured interviews with students after the completion of the class, to learn more about student experiences and work patterns.

It would also be interesting to know if students who spent more time on RStudio Cloud received higher or lower grades on their assignments, but as discussed in Section \ref{sec:assessment}, the IRB did not cover graded student work in that way. We do know the two sections did not have an overall difference in mean lab grade.

The results can also be used by instructors attempting to ballpark how many usage hours their classes may need over the course of a month or a semester. Students in the \texttt{tidyverse} section used an average of 13.5 hours per month, and students in the formula section used an average of 11.5 hours. These numbers can be used to make back-of-the-envelope calculations on how much RStudio Cloud would end up costing for a class of a particular size, or to make a decision about flat rate pricing, which could be most cost effective.

\subsection{Function use and repetition}\label{sec:numfunc}

\linespread{1}
\linespread{2}
\vspace{3mm}\setlength{\parindent}{15pt}

\linespread{1}
\linespread{2}
\vspace{3mm}\setlength{\parindent}{15pt}

Another source of incidental data is the set of RMarkdown documents, which allow us to study the number of functions students were exposed to in each section.

Cognitive load theory suggests showing students fewer functions and reusing them many times over the course of the semester would allow them to move some of the information from their working memory to their longterm memory \citep{lovettgreenhouse2000}. Because of this, it is ideal for instructors to minimize the number of functions they show students, and show each function at least twice \citep{mcnamaraetal2021}.

One argument against the use of the \texttt{tidyverse} in teaching is that it contains too many functions, and therefore is overwhelming for students. However, when teaching a course (particularly an introductory one) an instructor never shows all the functions in a package. Instead, it is crucial to identify just the most essential functions, ideally with consistent syntax.

\subsubsection{Counting functions}\label{counting-functions}

Since both sections relied on the use of RMarkdown documents, we have several sources of data on functions. First, the instructor prepared pre-lab documents with blanks, but also saved a `filled-in' copy after recording the accompanying video. The filled in version includes all the functions shown in the pre-lab video.

Students in each section were also given a ``All the R you need for intro stats'' cheatsheet at the beginning of the semester. These cheatsheets (one for formula and one for \texttt{tidyverse}) were modeled on the cheatsheet of a similar name accompanying the \texttt{mosaic} package \citep{pruimetal2017}. The cheatsheets aimed to include all code necessary for the entire semester, but were generated a priori.

These documents allowed for the use of automated methods to analyze the number of unique functions shown in each section, using the \texttt{getParseData()} function from the built-in \texttt{utils} package. The parsed data was filtered to only functions, and specifically functions with parenthetical notation (infix functions like \texttt{+} were not counted, nor was the pipe operator).

The parsing code was run on the filled-in pre-lab documents produced by the instructor at the end of recording the pre-lab videos. These filled-in documents represented all the functions the instructor demonstrated to the students. From this data, we found the formula section saw a total of 41 functions and the \texttt{tidyverse} section saw 52. For a list of the functions used in both sections, see Appendix.

It makes sense the \texttt{tidyverse} section would see a larger number of functions, because there are several elements of the \texttt{tidyverse} that require combining multiple functions to accomplish a single task. For example, to find a summary statistic like the mean, students in the \texttt{tidyverse} section needed to use code like \texttt{penguins\ \textbar{}\textgreater{}\ summarize(mean(flipper\_length\_mm))} (two nested functions), while the students in the formula section would write \texttt{mean(\textasciitilde{}flipper\_length\_mm,\ data\ =\ penguins)} (one function).

The parsing code was also run on the ``All the R you need for intro stats'' cheatsheets, to determine how many of the functions shown in class were included on the sheets produced ahead of time.

\linespread{1}
\linespread{2}
\vspace{3mm}\setlength{\parindent}{15pt}

The cheatsheets given to students at the beginning of the semester contained 35 functions for the formula section and 42 functions for the \texttt{tidyverse} section. There was an overlap of 18 functions between the two cheatsheets.

These numbers make it appear as if in the formula section the instructor showed all functions from the cheatsheet, and then a few additional functions. However, there were actually several functions on the cheatsheet that were never shown in the actual class, and many more functions that appeared in the class that did not make it onto the cheatsheet.

In the \texttt{tidyverse} section, there were 52 functions shown in class that did not appear on the cheatsheet, and only 42 function on the cheatsheet that were not discussed in class. In the formula section, there were also 41 functions shown in class that did not appear on the cheatsheet, as well as 35 functions on the cheatsheet that were not discussed in class. In both classes the majority of functions shown in class were on the cheatsheet.

These results helped the instructor in subsequent semesters, because she could better align the cheatsheet with the functions actually used throughout the semester.

\subsubsection{Function usage and repetition}\label{function-usage-and-repetition}

\linespread{1}
\linespread{2}
\vspace{3mm}\setlength{\parindent}{15pt}

Ideally, we believe students should see each function at least twice, to help them reinforce how it is used. Interestingly, although the \texttt{tidyverse} section saw more functions overall, it also had a higher proportion of reused functions. In the \texttt{tidyverse} section, 77\% of functions were shown more than once, and in the formula section, 63\% of functions were. In both sections, the majority of functions were shown more than once.

However, some functions were only shown once over the course of the entire semester. There were 15 functions shown only one time in the formula section, and 12 functions only shown once in the \texttt{tidyverse} section.

\linespread{1}
\linespread{2}
\vspace{3mm}\setlength{\parindent}{15pt}

\linespread{1}
\linespread{2}
\vspace{3mm}\setlength{\parindent}{15pt}

\linespread{1}
\begin{figure}

{\centering \includegraphics[width=0.8\linewidth]{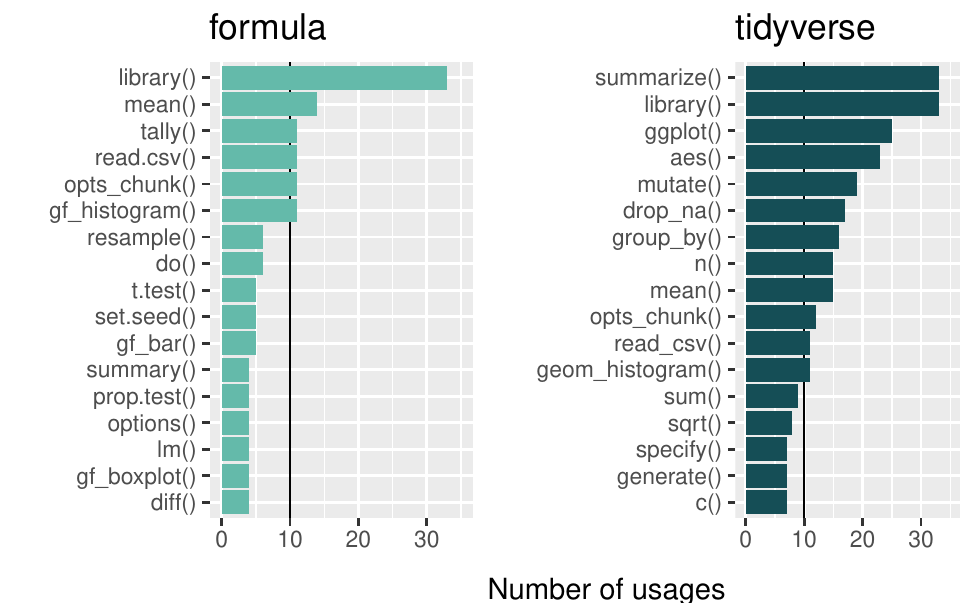} 

}

\caption{Top 15 functions in each section. The formula section saw a total of 41 functions and the \texttt{tidyverse} section saw 52. Notice that the \texttt{tidyverse} section has many more functions repeated more than 10 times during the semester. Overall, 63\% of formula functions were repeated, and 77\% of \texttt{tidyverse} functions were.}\label{fig:common-functions}
\end{figure}\linespread{2}
\vspace{3mm}\setlength{\parindent}{15pt}

In Figure \ref{fig:common-functions} we can see the 15 most-used functions in each section. The full list of functions and their usage data is available in the GitHub repo for this paper (\url{https://github.com/AmeliaMN/ComparingSyntaxForModeling}). Both sections used the \texttt{library()} command three times per lab, so those counts are identical between the sections. Similarly, both sections read in data once per lab, so the count for \texttt{read\_csv()} in the \texttt{tidyverse} section is the same as the count for \texttt{read.csv()} in the formula section. The difference in rank for the data-reading code (\texttt{read.csv()} was one of the third-most-common functions for the formula section, but \texttt{read\_csv()} was one of the ninth-most-common for the \texttt{tidyverse} section) makes it clear there are differences in the pattern of function usage between the sections.

In the formula section, the summary statistic \texttt{mean()} was the second-ranked function, used times. There were four functions used times-- \texttt{tally()}, \texttt{opts\_chunk()} (a function in the \texttt{knitr} options included in the top of each RMarkdown document) , the aforementioned \texttt{read.csv()} and \texttt{gf\_histogram()}.

In the \texttt{tidyverse} section, \texttt{mean()} was also used frequently ( times), but it is nowhere near the top of the list. Instead, more general functions like \texttt{summarize()} (used times) and \texttt{ggplot()} ( times) top the list.

So, while the formula section used fewer functions overall, they were also not repeated as much. While the \texttt{tidyverse} section saw more functions, they saw some functions many times. It makes sense that the repeated functions in the \texttt{tidyverse} section would be used more frequently, as tasks in the \texttt{tidyverse} tend to use common functions for plotting and summary statistics, combined with additional specialty functions.

For example, students in both sections saw how to make a barchart, boxplot, histogram, and scatterplot, but in the formula section they used standalone functions like \texttt{gf\_boxplot()} whereas in the \texttt{tidyverse} section they needed to start with \texttt{ggplot()} and add on a \texttt{geom\_*()} function like \texttt{geom\_boxplot()}, while specifying the \texttt{aes()}thetic values somewhere.

\subsubsection{Shared functions}\label{shared-functions}

Interestingly, there was some overlap in the functions students saw in both sections. In fact, there were 21 function names both sections saw, including helper functions like \texttt{library()}, \texttt{set.seed()}, and \texttt{opts\_chunk()}, modeling-related functions like \texttt{aov()}, \texttt{lm()}, \texttt{summary()} and \texttt{predict()}, and statistics like \texttt{mean()}, \texttt{sd()}, and \texttt{cor()}.

\linespread{1}
\linespread{2}
\vspace{3mm}\setlength{\parindent}{15pt}

However, while both sections saw \texttt{mean()}, the \texttt{tidyverse} section used the \textcolor{grey}{base::}\texttt{mean()} function and the formula section saw the \texttt{mosaic} masked version of mean, \textcolor{grey}{mosaic::}\texttt{mean()}. If we count the masked version of functions separately, the overlap between sections reduces to 16. If we consider masked versions of functions different, students in the formula section saw 25 functions apart from the set both sections saw, while the \texttt{tidyverse} section saw 36 functions beyond the common set. Again, this makes sense based on the way \texttt{tidyverse} code is structured.

On the subject of modeling, the instructor chose not to introduce the \texttt{tidymodels} package to the \texttt{tidyverse} section. \texttt{tidymodels} provides a consistent interface to modeling that better aligns with the \texttt{tidyverse}, using a data-first syntax that allows the use of the pipe \citet{kuhnetal2022}. However, it still uses the \texttt{\textasciitilde{}} and \texttt{data\ =} syntax when fitting a linear model, which means it still uses both syntaxes under discussion in this work. In order to reduce the number of \texttt{library()} calls for students, the instructor chose not to teach with \texttt{tidymodels} in this course.

\subsubsection{Pedagogical implications}\label{pedagogical-implications}

Overall, neither section appeared to expose students to an overwhelming number of functions. The \texttt{tidyverse} section saw 52 as compared to the formula section's 41 functions, but that difference does not feel practically significant, particularly considering the way in which \texttt{tidyverse} operations often use helper functions. While the \texttt{tidyverse} section saw more functions, their most common functions were also repeated more. This additional repetition may have helped students move those common functions out of working memory to longterm memory.

As noted in Section \ref{sec:yt}, the instructional materials for the \texttt{tidyverse} section were slightly longer than those for the formula section, but the difference was slight (videos 2 minutes longer or 9\% longer). Even though more functions were used, the explanation did not take significantly longer.

The comparison also underscores the fact that while instructors may say they are teaching \texttt{tidyverse} or formula syntax, they are ultimately teaching R. Both sections saw 16 common functions, most of them from base R (the \texttt{base}, \texttt{utils} or \texttt{stats} packages that come standard with the language).

The practice of analyzing the number of functions shown over the course of the semester was eye-opening. It has already provided valuable information for the instructor, as she was able to remove some functions only shown once, and better match the cheatsheets to what is shown throughout the semester in subsequent courses. The list of functions provided in the Appendix can also serve as a starting point for other instructors as they work to produce curricular materials for introductory statistics classes in R.

\subsection{Divergent labs}\label{sec:diflabs}

So far, we have primarily focused on similarities between the experiences of the two sections. Some data sources have shown slight differences between the syntaxes, but the primary source of information on how they vary is qualitative experience. The function counting exercise began to point toward these qualitative differences.

From the instructor's experience, there are three primary places where the experience diverged.

\subsubsection{Summary statistics for categorical variables}\label{summary-statistics-for-categorical-variables}

In the sequence of topics for this course, the first lab was focused on ``describing data.'' In both labs, the instructor showed only five lines of code-- \texttt{knitr} options that were already in the document and explained only briefly, three \texttt{library()} calls, and a single line to load in the dataset. From there, all of the content was looking at the data using the RStudio interface, without any coding. This allowed for a gradual introduction to the tools. The code for the first lab was extremely simple, but students learned how to log in to RStudio Cloud, run code, and knit an RMarkdown document.

The second lab of the semester focused on exploratory data analysis of one and two categorical variables. Because this was the first lab where students needed to apply multiple functions to their own data, it was bound to be longer and more challenging. However, there were additional complications introduced in the \texttt{tidyverse} lab.

In this week (week 3 of the semester), the labs were of particularly different lengths. For the formula section the RMarkdown document was 89 lines long, and the two videos totaled 28 minutes. The RMarkdown document for the \texttt{tidyverse} section was 125 lines long, and the videos totaled 35 minutes. There is a clear reason why.

\linespread{1}
\linespread{2}
\vspace{3mm}\setlength{\parindent}{15pt}

In the formula section, students found frequency tables and relative frequency tables with code as in \ref{tally-ex1} and \ref{tally-ex2}.

\linespread{1}
\linespread{2}
\vspace{3mm}\setlength{\parindent}{15pt}

\linespread{1}

\begin{Shaded}
\begin{Highlighting}[]
\FunctionTok{tally}\NormalTok{(}\SpecialCharTok{\textasciitilde{}}\NormalTok{island, }\AttributeTok{data =}\NormalTok{ penguins)}
\FunctionTok{tally}\NormalTok{(}\SpecialCharTok{\textasciitilde{}}\NormalTok{island, }\AttributeTok{data =}\NormalTok{ penguins, }\AttributeTok{format =} \StringTok{"percent"}\NormalTok{)}
\FunctionTok{tally}\NormalTok{(species }\SpecialCharTok{\textasciitilde{}}\NormalTok{ island, }\AttributeTok{data =}\NormalTok{ penguins)}
\end{Highlighting}
\end{Shaded}

\captionof{chunk}{Making tables of one and two categorical variables using the formula syntax and \texttt{tally()} from the \texttt{mosaic} package.}

\label{tally-ex1}
\linespread{2}
\vspace{3mm}\setlength{\parindent}{15pt}

\linespread{1}

\begin{Shaded}
\begin{Highlighting}[]
\FunctionTok{tally}\NormalTok{(species }\SpecialCharTok{\textasciitilde{}}\NormalTok{ island, }\AttributeTok{data =}\NormalTok{ penguins, }\AttributeTok{format =} \StringTok{"percent"}\NormalTok{)}
\end{Highlighting}
\end{Shaded}

\begin{verbatim}
           island
species        Biscoe     Dream Torgersen
  Adelie     26.19048  45.16129 100.00000
  Chinstrap   0.00000  54.83871   0.00000
  Gentoo     73.80952   0.00000   0.00000
\end{verbatim}

\captionof{chunk}{Making a table of two categorical variables using the formula syntax and \texttt{tally()} function, along with the percent option.}

\label{tally-ex2}
\linespread{2}
\vspace{3mm}\setlength{\parindent}{15pt}

The \texttt{tally()} function produces a familiar-looking two-way table, which took very little explanation, other than to show how reversing the variables in the formula led to different percentages, as is seen in \ref{tally-ex3}. Compare \ref{tally-ex2} and \ref{tally-ex3} to see the effect of swapping the order of variables.

\linespread{1}

\begin{Shaded}
\begin{Highlighting}[]
\FunctionTok{tally}\NormalTok{(island }\SpecialCharTok{\textasciitilde{}}\NormalTok{ species, }\AttributeTok{data =}\NormalTok{ penguins, }\AttributeTok{format =} \StringTok{"percent"}\NormalTok{)}
\end{Highlighting}
\end{Shaded}

\begin{verbatim}
           species
island         Adelie Chinstrap    Gentoo
  Biscoe     28.94737   0.00000 100.00000
  Dream      36.84211 100.00000   0.00000
  Torgersen  34.21053   0.00000   0.00000
\end{verbatim}

\captionof{chunk}{Making a table of two categorical variables using the formula syntax and \texttt{tally()} function, with variables swapped.}

\label{tally-ex3}
\linespread{2}
\vspace{3mm}\setlength{\parindent}{15pt}

\linespread{1}
\linespread{2}
\vspace{3mm}\setlength{\parindent}{15pt}

However, in the \texttt{tidyverse} section, both the code and output took longer to explain. Initial summary statistics for categorical variables are computed in \ref{tidy-tally1}, while the tidy version of a relative frequency table is shown in \ref{tidy-tally2}.

\linespread{1}

\begin{Shaded}
\begin{Highlighting}[]
\NormalTok{penguins }\SpecialCharTok{|\textgreater{}}
  \FunctionTok{group\_by}\NormalTok{(island) }\SpecialCharTok{|\textgreater{}}
  \FunctionTok{summarize}\NormalTok{(}\AttributeTok{n =} \FunctionTok{n}\NormalTok{())}

\NormalTok{penguins }\SpecialCharTok{|\textgreater{}}
  \FunctionTok{group\_by}\NormalTok{(island) }\SpecialCharTok{|\textgreater{}}
  \FunctionTok{summarize}\NormalTok{(}\AttributeTok{n =} \FunctionTok{n}\NormalTok{()) }\SpecialCharTok{|\textgreater{}}
  \FunctionTok{mutate}\NormalTok{(}\AttributeTok{prop =}\NormalTok{ n }\SpecialCharTok{/} \FunctionTok{sum}\NormalTok{(n))}

\NormalTok{penguins }\SpecialCharTok{|\textgreater{}}
  \FunctionTok{group\_by}\NormalTok{(island, species) }\SpecialCharTok{|\textgreater{}}
  \FunctionTok{summarize}\NormalTok{(}\AttributeTok{n =} \FunctionTok{n}\NormalTok{())}
\end{Highlighting}
\end{Shaded}

\captionof{chunk}{Computing summary statistics for one and two categorical variables in the \texttt{tidyverse} syntax. These functions come from the \texttt{dplyr} package of the \texttt{tidyverse}.}

\label{tidy-tally1}
\linespread{2}
\vspace{3mm}\setlength{\parindent}{15pt}

\linespread{1}

\begin{Shaded}
\begin{Highlighting}[]
\NormalTok{penguins }\SpecialCharTok{|\textgreater{}}
  \FunctionTok{group\_by}\NormalTok{(island, species) }\SpecialCharTok{|\textgreater{}}
  \FunctionTok{summarize}\NormalTok{(}\AttributeTok{n =} \FunctionTok{n}\NormalTok{()) }\SpecialCharTok{|\textgreater{}}
  \FunctionTok{mutate}\NormalTok{(}\AttributeTok{prop =}\NormalTok{ n }\SpecialCharTok{/} \FunctionTok{sum}\NormalTok{(n))}
\end{Highlighting}
\end{Shaded}

\begin{verbatim}
# A tibble: 5 x 4
# Groups:   island [3]
  island    species       n  prop
  <fct>     <fct>     <int> <dbl>
1 Biscoe    Adelie       44 0.262
2 Biscoe    Gentoo      124 0.738
3 Dream     Adelie       56 0.452
4 Dream     Chinstrap    68 0.548
5 Torgersen Adelie       52 1    
\end{verbatim}

\captionof{chunk}{Computing summary statistics for two categorical variables in the \texttt{tidyverse} syntax.}

\label{tidy-tally2}
\linespread{2}
\vspace{3mm}\setlength{\parindent}{15pt}

Again, reversing the order of the variables (this time, inside the \texttt{group\_by()} function) changed the percentages, but it was more difficult to determine how the percentages added up, because the data was in long format, rather than wide format. Compare \ref{tidy-tally2} and \ref{tidy-tally3} to see the effect of swapping the order of variables.

\linespread{1}

\begin{Shaded}
\begin{Highlighting}[]
\NormalTok{penguins }\SpecialCharTok{|\textgreater{}}
  \FunctionTok{group\_by}\NormalTok{(species, island) }\SpecialCharTok{|\textgreater{}}
  \FunctionTok{summarize}\NormalTok{(}\AttributeTok{n =} \FunctionTok{n}\NormalTok{()) }\SpecialCharTok{|\textgreater{}}
  \FunctionTok{mutate}\NormalTok{(}\AttributeTok{prop =}\NormalTok{ n }\SpecialCharTok{/} \FunctionTok{sum}\NormalTok{(n))}
\end{Highlighting}
\end{Shaded}

\begin{verbatim}
# A tibble: 5 x 4
# Groups:   species [3]
  species   island        n  prop
  <fct>     <fct>     <int> <dbl>
1 Adelie    Biscoe       44 0.289
2 Adelie    Dream        56 0.368
3 Adelie    Torgersen    52 0.342
4 Chinstrap Dream        68 1    
5 Gentoo    Biscoe      124 1    
\end{verbatim}

\captionof{chunk}{Computing summary statistics for two categorical variables in the \texttt{tidyverse} syntax, with variables swapped.}

\label{tidy-tally3}
\linespread{2}
\vspace{3mm}\setlength{\parindent}{15pt}

\subsubsection{Summary statistics for quantitative variables}\label{summary-statistics-for-quantitative-variables}

As discussed in Section \ref{sec:numfunc}, one reason why the \texttt{tidyverse} section saw more functions than the formula section is the way summary statistics are computed in the \texttt{tidyverse}.

In formula syntax, summary statistics were always a single function call, like \texttt{mean()}, \texttt{sd()} or \texttt{favstats()}, whereas for the \texttt{tidyverse} section those summary functions needed to be wrapped within the \texttt{summarize()} function.

In the lab focused on exploratory data analysis for quantitative variables, the formula students found the mean as shown in \ref{formulamean}.

\linespread{1}
\linespread{2}
\vspace{3mm}\setlength{\parindent}{15pt}

\linespread{1}

\begin{Shaded}
\begin{Highlighting}[]
\FunctionTok{mean}\NormalTok{(}\SpecialCharTok{\textasciitilde{}}\NormalTok{bill\_length\_mm, }\AttributeTok{data =}\NormalTok{ penguins, }\AttributeTok{na.rm =} \ConstantTok{TRUE}\NormalTok{)}
\end{Highlighting}
\end{Shaded}

\captionof{chunk}{Using the \texttt{mosaic} \texttt{mean()} function to compute the mean. \texttt{mosaic} masks the base version of \texttt{mean()}, and makes it formula-aware.}

\label{formulamean}
\linespread{2}
\vspace{3mm}\setlength{\parindent}{15pt}

\linespread{1}
\linespread{2}
\vspace{3mm}\setlength{\parindent}{15pt}

In the \texttt{tidyverse} section, they found the mean as shown in \ref{meantidy}

\linespread{1}

\begin{Shaded}
\begin{Highlighting}[]
\NormalTok{penguins }\SpecialCharTok{|\textgreater{}}
  \FunctionTok{drop\_na}\NormalTok{(bill\_length\_mm) }\SpecialCharTok{|\textgreater{}}
  \FunctionTok{summarize}\NormalTok{(}\FunctionTok{mean}\NormalTok{(bill\_length\_mm))}
\end{Highlighting}
\end{Shaded}

\captionof{chunk}{Finding the mean using \texttt{tidyverse} syntax. The \texttt{drop\_na()} and \texttt{summarize()} functions are from the \texttt{dplyr} package, \texttt{mean()} comes from base R.}

\label{meantidy}
\linespread{2}
\vspace{3mm}\setlength{\parindent}{15pt}

The formula approach has the benefit of being a single line, but it relies on the use of \texttt{na.rm\ =\ TRUE}, which students found hard to remember. The instructor also chose to show how to change the default \texttt{NA} action by using \texttt{options(na.rm\ =\ TRUE)}. This was more popular with students, because they didn't have to add that argument to their summary statistic functions, but became confusing in the subsequent lab when the \texttt{cor()} function required the \texttt{use\ =\ "complete.obs"} argument.

Dealing with missing values was more verbose in the \texttt{tidyverse} section, but also was easier to explain to students. And, the \texttt{drop\_na()} approach could be easily extended for the subsequent lab, removing the need to talk about \texttt{use\ =\ "complete.obs"}.

After doing single summary statistics for a quantitative variable, the lab asked students to complete a five number summary. In the formula lab, students found the five number summary as shown in \ref{favstats}.

\linespread{1}
\linespread{2}
\vspace{3mm}\setlength{\parindent}{15pt}

\linespread{1}

\begin{Shaded}
\begin{Highlighting}[]
\FunctionTok{favstats}\NormalTok{(}\SpecialCharTok{\textasciitilde{}}\NormalTok{bill\_length\_mm, }\AttributeTok{data =}\NormalTok{ penguins)}
\end{Highlighting}
\end{Shaded}

\captionof{chunk}{The \texttt{mosaic} \texttt{favstats()} function provides many common summary statistics for one quantitative variable. The \texttt{favstats()} function automatically drops missing values.}

\label{favstats}
\linespread{2}
\vspace{3mm}\setlength{\parindent}{15pt}

\linespread{1}
\linespread{2}
\vspace{3mm}\setlength{\parindent}{15pt}

This approach deals with missing values as part of the standard output, which is appealing to students but also provides an inconsistency that is hard to explain.

In the \texttt{tidyverse} section, the instructor chose to show two approaches. Both approaches are in \ref{tidy-summary1}, and both needed to include \texttt{drop\_na()} to deal with missing values. Past those similarities, the approaches are divergent.

\linespread{1}

\begin{Shaded}
\begin{Highlighting}[]
\NormalTok{penguins }\SpecialCharTok{|\textgreater{}}
  \FunctionTok{drop\_na}\NormalTok{(bill\_length\_mm) }\SpecialCharTok{|\textgreater{}}
  \FunctionTok{summarize}\NormalTok{(}
    \AttributeTok{min =} \FunctionTok{min}\NormalTok{(bill\_length\_mm),}
    \AttributeTok{lower\_hinge =} \FunctionTok{quantile}\NormalTok{(bill\_length\_mm, .}\DecValTok{25}\NormalTok{),}
    \AttributeTok{median =} \FunctionTok{median}\NormalTok{(bill\_length\_mm),}
    \AttributeTok{upper\_hinge =} \FunctionTok{quantile}\NormalTok{(bill\_length\_mm, .}\DecValTok{75}\NormalTok{),}
    \AttributeTok{max =} \FunctionTok{max}\NormalTok{(bill\_length\_mm)}
\NormalTok{  )}

\NormalTok{penguins }\SpecialCharTok{|\textgreater{}}
  \FunctionTok{drop\_na}\NormalTok{(bill\_length\_mm) }\SpecialCharTok{|\textgreater{}}
  \FunctionTok{pull}\NormalTok{(bill\_length\_mm) }\SpecialCharTok{|\textgreater{}}
  \FunctionTok{fivenum}\NormalTok{()}
\end{Highlighting}
\end{Shaded}

\captionof{chunk}{Two approaches for doing summary statistics of one quantitative variable in \texttt{tidyverse} syntax. The first is quite verbose, the second is more compact but introduces a function never seen again. Both approaches use functions from the \texttt{dplyr} package.}

\label{tidy-summary1}
\linespread{2}
\vspace{3mm}\setlength{\parindent}{15pt}

It would have been preferable for the instructor to choose a single solution to present to students, but she was faced with a dilemma. The first \texttt{tidyverse} approach is very verbose, but it follows nicely from other summary statistics students had already seen, just adding a few more functions like \texttt{min()}, \texttt{max()}, and \texttt{quantile()}. The second solution is more concise, but it introduces the \texttt{pull()} function, which was never used again in the course. This brings back the consideration of how many times students will see the same function.

Another \texttt{tidyverse} option more in line with the formula \texttt{favstats()} function would have been to use the \texttt{skim()} function from the \texttt{skimr} package \citep{mcnamaraetal2018}. However, the instructor was trying to keep the number of \texttt{library()} calls at the beginning of the document consistent over the semester, and as low as possible. From the perspective of cognitive load, however, introducing a single new R package with a function that could be used repeatedly might have been better than introducing a function in a familiar package that was only seen once in the semester.

\subsubsection{Inference for two categorical variables}\label{inference-for-two-categorical-variables}

In week 10, the topic was inference for two categorical variables. In this week, students returned to the summary statistics from week 3, and extended their work to inferential statistics.

Similar to the week with descriptive statistics, this week's labs diverged in length. The formula section's RMarkdown document was 89 lines long, and the videos totaled 19 minutes. That same week, the \texttt{tidyverse} RMarkdown document was 119 lines long, and the videos totaled 27 minutes.

The explanation for the varying time is similar, as well. Week 10 focused on inference for two samples; that is, inference for a difference of proportions or a difference of means. While a difference of means makes it fairly easy to know which variable should go where (the quantitative variable is the response variable to take the mean of, and the categorical variable is the explanatory variable splitting it), with a difference of two proportions the concept comes back to thinking about two-way tables. Again, the \texttt{tidyverse} presentation of a ``two-way table'' made this more difficult to conceptualize.

In the formula section, students saw code like that in \ref{formula-prop}.

\linespread{1}
\linespread{2}
\vspace{3mm}\setlength{\parindent}{15pt}

\linespread{1}

\begin{Shaded}
\begin{Highlighting}[]
\FunctionTok{tally}\NormalTok{(island }\SpecialCharTok{\textasciitilde{}}\NormalTok{ sex, }\AttributeTok{data =}\NormalTok{ penguins, }\AttributeTok{format =} \StringTok{"proportion"}\NormalTok{)}
\end{Highlighting}
\end{Shaded}

\begin{verbatim}
        sex
island      female      male
  Biscoe 0.5673759 0.5724138
  Dream  0.4326241 0.4275862
\end{verbatim}

\begin{Shaded}
\begin{Highlighting}[]
\FunctionTok{prop.test}\NormalTok{(island }\SpecialCharTok{\textasciitilde{}}\NormalTok{ sex, }\AttributeTok{data =}\NormalTok{ penguins, }\AttributeTok{success =} \StringTok{"Biscoe"}\NormalTok{)}
\end{Highlighting}
\end{Shaded}

\begin{verbatim}

    2-sample test for equality of proportions with continuity correction

data:  tally(island ~ sex)
X-squared = 2.8876e-30, df = 1, p-value = 1
alternative hypothesis: two.sided
95 percent confidence interval:
 -0.1248439  0.1147680
sample estimates:
   prop 1    prop 2 
0.5673759 0.5724138 
\end{verbatim}

\captionof{chunk}{Making a two-way table and performing inference for a difference of proportions using the formula syntax. In order for this code to run as-is, the Torgerson island has to be removed so there are just two categories in the variable    exttt{island}.}

\label{formula-prop}
\linespread{2}
\vspace{3mm}\setlength{\parindent}{15pt}

The code for finding the point estimate using \texttt{tally()} is quite similar to the code for performing inference using \texttt{prop.test()}. And, the output from \texttt{prop.test()} includes the sample estimates, for comparison with the numbers seen in the initial table. This allows students to verify they are comparing the proportions they intended to.

In the \texttt{tidyverse} section, the code was not as consistent. Students in this section saw code like that shown in \ref{tidy-prop}.

\linespread{1}
\linespread{2}
\vspace{3mm}\setlength{\parindent}{15pt}

\linespread{1}

\begin{Shaded}
\begin{Highlighting}[]
\NormalTok{penguins }\SpecialCharTok{|\textgreater{}}
  \FunctionTok{group\_by}\NormalTok{(sex, island) }\SpecialCharTok{|\textgreater{}}
  \FunctionTok{summarize}\NormalTok{(}\AttributeTok{n =} \FunctionTok{n}\NormalTok{()) }\SpecialCharTok{|\textgreater{}}
  \FunctionTok{mutate}\NormalTok{(}\AttributeTok{prop =}\NormalTok{ n }\SpecialCharTok{/} \FunctionTok{sum}\NormalTok{(n))}
\end{Highlighting}
\end{Shaded}

\begin{verbatim}
# A tibble: 4 x 4
# Groups:   sex [2]
  sex    island     n  prop
  <fct>  <fct>  <int> <dbl>
1 female Biscoe    80 0.567
2 female Dream     61 0.433
3 male   Biscoe    83 0.572
4 male   Dream     62 0.428
\end{verbatim}

\begin{Shaded}
\begin{Highlighting}[]
\FunctionTok{library}\NormalTok{(infer)}
\NormalTok{penguins }\SpecialCharTok{|\textgreater{}}
  \FunctionTok{prop\_test}\NormalTok{(}
    \AttributeTok{response =}\NormalTok{ island,}
    \AttributeTok{explanatory =}\NormalTok{ sex,}
    \AttributeTok{alternative =} \StringTok{"two{-}sided"}\NormalTok{,}
    \AttributeTok{order =} \FunctionTok{c}\NormalTok{(}\StringTok{"female"}\NormalTok{, }\StringTok{"male"}\NormalTok{)}
\NormalTok{  )}
\end{Highlighting}
\end{Shaded}

\begin{verbatim}
# A tibble: 1 x 6
  statistic chisq_df p_value alternative lower_ci upper_ci
      <dbl>    <dbl>   <dbl> <chr>          <dbl>    <dbl>
1  1.78e-30        1    1.00 two.sided     -0.127    0.117
\end{verbatim}

\captionof{chunk}{Making a `two-way table' and performing inference for a difference of proportions using the \texttt{tidyverse} syntax. Again, the Torgerson island data has been removed beforehand.}

\label{tidy-prop}
\linespread{2}
\vspace{3mm}\setlength{\parindent}{15pt}

In \texttt{tidyverse} syntax the code for finding the point estimate (\texttt{dplyr}'s \texttt{group\_by()}, \texttt{summarize()} and then \texttt{mutate()}) is quite different from the code performing the inference (the \texttt{infer} \texttt{prop\_test()} function). And, the output from the inferential \texttt{prop\_test()} function does not include the sample estimates, which makes it harder to determine if the code was correct.

These discrepancies made it take longer to explain code in the \texttt{tidyverse} section for these topics.

\section{Discussion}\label{sec:discussion}

This semester-long, head-to-head comparison of two sections of introductory statistics labs provides data comparing two popular R coding styles, the formula syntax and the \texttt{tidyverse} syntax. Materials for the \texttt{tidyverse} section tended to be longer in lines of code (likely because of the convention of linebreaks after \texttt{\textbar{}\textgreater{}}), and had slightly longer associated YouTube videos, although this difference was minimal (approximately 2 minutes longer, on average)

Pre- and post-survey analysis showed limited differences between the two sections, although analysis of other incidental data, including YouTube and RStudio Cloud data presented interesting distinctions.

Students in the \texttt{tidyverse} section watched a smaller proportion of the weekly pre-lab videos than students in the formula section, but seemed to spend more time computing on RStudio. Conversely, students in the formula section watched a larger proportion of the pre-lab videos each week, but generally spent less time computing each month.

These two insights are slightly contradictory-- perhaps the formula section students found the concepts more complex as they were watching the videos, but then had an easier time applying them as they worked on the lab.

\linespread{1}
\linespread{2}
\vspace{3mm}\setlength{\parindent}{15pt}

The \texttt{tidyverse} section exposed students to 52, compared to the 41 functions shown in the formula section. Both labs focused on a relatively small number of functions. Because there were 12 labs in the semester, this averages out to approximately 4 functions per lab for the \texttt{tidyverse} section compared to an average 3 functions shown in the formula section. The \texttt{tidyverse} section saw more unique functions, but both sections were limited to a small vocabulary of functions for the semester.

Beyond the difference in number of functions shown, there were also differences in the amount of repetition functions saw in the two sections. In the \texttt{tidyverse} section, there were more functions shown more than once, and a larger proportion of functions were repeated. The most common functions in the \texttt{tidyverse} section were also shown many more times than the most common functions in the formula section.

Regardless of the syntax they choose, we recommend instructors attempt to reduce the number of functions they expose students to over the course of a semester, particularly in an introductory class, and repeat them as much as possible. This will help reduce cognitive load. If optimizing for the minimum number of functions, the formula syntax may be the best solution. If trying to maximize repetition, the \texttt{tidyverse} syntax appears better.

The exercise of counting R functions in existing materials, using the \texttt{getParseData()} function, is one we recommend all instructors attempt, particularly before re-teaching a course. It can be eye-opening to discover how many functions are shown to students, and which functions are only used once.

Overall, the experience of teaching the two syntaxes was relatively similar. The R packages that have been developed to support student learning in introductory courses are robust, and made it possible to provide positive experiences to both groups.

One set of topics where the experience diverged significantly was dealing with two categorical variables, both in terms of summary statistics and inferential statistics. Students seem to struggle with conceptualizing of relationships between two categorical variables in general (perhaps related to issues considering conditional probability), and adding code into the mix made it no easier.

These topics were harder to do and explain in the \texttt{tidyverse}. When computing summary statistics, the long format of the output was not as familiar as the two-way table produced by the \texttt{mosaic} function \texttt{tally()}. The challenges with inferential statistics followed on from those in summary statistics, with the added challenge that the function for inference in the \texttt{tidyverse} (\texttt{infer}'s \texttt{prop\_test()}) is not as parallel with those for summary statistics as are the corresponding functions in formula syntax. However, the authors of the \texttt{infer} package continue to develop the package, and future versions of the \texttt{prop\_test()} and other similar functions may solve this issue.

Summary statistics for quantitative variables suffered from a different set of challenges. In the formula section, dealing with missing values was inconsistent (sometimes \texttt{na.rm\ =\ TRUE}, sometimes \texttt{use\ =\ "complete.obs"}), but finding a five number summary with \texttt{favstats()} was straightforward. In the \texttt{tidyverse} section, dealing with missing values using \texttt{drop\_na()} made that process very consistent, but finding a five number summary was verbose. This challenge could be overcome by the use of the \texttt{skimr} package for summary statistics.

Overall, most issues with each syntax could be solved with better preparation, including more consideration of consistency. There is no clear answer to which syntax is `best' but these results and discussion may help instructors choose the syntax most appropriate for their course and student population.

There is interesting further work that could be considered. A cross-over design where students saw one syntax for the first half of the semester and the other for the second half would allow for better comparisons. However, there are a few caveats here.

First, anecdotal evidence from many instructors suggests it is best for students to see only one syntax over the course of the semester. The other challenge is the formula syntax tends to seep (albeit only minorly) into the \texttt{tidyverse} section. For example, when doing linear regression both sections saw the \texttt{lm(y\textasciitilde{}x,\ data\ =\ data)} formula syntax, because the instructor chose not to introduce the \texttt{tidymodels} package. If a cross-over design used the existing materials from these classes, just swapping the final few weeks, students in the formula section would likely see more that was familiar to them than students in the \texttt{tidyverse} section. This could potentially be remedied by the inclusion of \texttt{tidymodels} for things like linear regression, because \texttt{tidymodels} offers regression modeling that is more consistent with \texttt{tidyverse} syntax

In fact, the \texttt{tidyverse} students almost \emph{did} have a cross-over design, because they were exposed to the standard \texttt{lm()} function, which used the formula syntax. This may be why the number of hours of compute time for the \texttt{tidyverse} section remained consistent from November to December while the formula section's hours of compute time decreased.

Another follow-up study that would be interesting to complete would look at student success in subsequent courses. Because \texttt{tidyverse} syntax is frequently used for higher-level courses, students who were in the \texttt{tidyverse} section may have an easier time in those later courses. However, most students in the classes under consideration will not go on to take further statistics courses. So the takeaways about syntax choice may vary depending on the student population to which they will be applied.

We hope this work helps answer some initial questions about the impact of R syntax on teaching introductory statistics, while also raising questions for future study. While some aspects of the analysis from these classes suggest the formula syntax is simpler for students to learn and use, there are still many course scenarios for which we believe the \texttt{tidyverse} syntax is the most appropriate choice. While formula syntax can be used throughout an entire semester of introductory statistics, it does not offer functionality for tasks like data wrangling. This means students who will go on to additional statistics or data science classes may be better served by an early introduction to \texttt{tidyverse}. However, in order to determine this conclusively, additional study would be needed.

No matter which syntax an instructor chooses, it appears possible to reduce the cognitive load of coding in R by limiting the number of functions shown in a semester, and provide students with a positive learning experience.

\subsection{Resources, packages, and data availability}\label{resources-packages-and-data-availability}

All pedagogical materials used for the course under discussion are available on GitHub and are Creative Commons licensed, so they can be used or remixed by anyone who wants (\url{https://github.com/AmeliaMN/STAT220-labs}). All code and anonymized data from this paper is also available on GitHub, for reproducibility (\url{https://github.com/AmeliaMN/ComparingSyntaxForModeling}). The same information is mirrored on the Open Science Framework (\url{doi.org/10.17605/OSF.IO/Y5TFV}). Data analysis was performed in R, and the paper was written in RMarkdown. The categorical color palette was chosen using Colorgorical \citep{gramazioetal2017}, and colors for the Likert scale plot are from ColorBrewer \citep{harrowerbrewer2003}. Example data used throughout the paper is from \texttt{palmerpenguins} \citep{horstetal2020}. Code from the formula section uses the R packages loaded in that course, \texttt{mosaic} and \texttt{ggformula} (\texttt{ggformula} is now loaded automatically with \texttt{mosaic}) \citep{pruimetal2017, kaplanpruim2020}. Code from the \texttt{tidyverse} section uses the packages from that course, \texttt{tidyverse} and \texttt{infer} \citep{wickhametal2019, brayetal2021}.

\section{Acknowledgements}\label{acknowledgements}

Thanks to Sean Kross for his guidance about parsing R function data and Nick Horton for his useful comments on the paper overall. Thanks also to the anonymous reviewers, whose comments have strengthened this paper considerably.

\clearpage

\appendix

\section{Functions used}\label{sec:functions}

\noindent

\begin{singlespace}
\begin{multicols*}{3}
\noindent
\linespread{1}
\noindent \newline\texttt{\textcolor{grey}{stats::}aov()} \newline\texttt{\textcolor{grey}{base::}data.frame()} \newline\texttt{\textcolor{grey}{dplyr::}filter()} \newline\texttt{\textcolor{grey}{base::}library()} \newline\texttt{\textcolor{grey}{stats::}lm()} \newline\texttt{\textcolor{grey}{base::}nrow()} \newline\texttt{\textcolor{grey}{stats::}pnorm()} \newline\texttt{\textcolor{grey}{stats::}predict()} \newline\texttt{\textcolor{grey}{stats::}pt()} \newline\texttt{\textcolor{grey}{stats::}qnorm()} \newline\texttt{\textcolor{grey}{stats::}qt()} \newline\texttt{\textcolor{grey}{knitr::}opts\_chunk()} \newline\texttt{\textcolor{grey}{base::}set.seed()} \newline\texttt{\textcolor{grey}{base::}sqrt()} \newline\texttt{\textcolor{grey}{base::}summary()} \newline\texttt{\textcolor{grey}{stats::}TukeyHSD()} \newline\linespread{2}
\vspace{3mm}\setlength{\parindent}{15pt}

\noindent
Used in both sections.
\columnbreak

\noindent
\linespread{1}
\noindent \newline\texttt{\textcolor{grey}{stats::}chisq.test()} \newline\texttt{\textcolor{grey}{mosaic::}confint()} \newline\texttt{\textcolor{grey}{mosaic::}cor()} \newline\texttt{\textcolor{grey}{base::}diff()} \newline\texttt{\textcolor{grey}{mosaic::}do()} \newline\texttt{\textcolor{grey}{mosaic::}factorize()} \newline\texttt{\textcolor{grey}{mosaic::}favstats()} \newline\texttt{\textcolor{grey}{mosaic::}fivenum()} \newline\texttt{\textcolor{grey}{ggformula::}gf\_bar()} \newline\texttt{\textcolor{grey}{ggformula::}gf\_boxplot()} \newline\texttt{\textcolor{grey}{ggformula::}gf\_density()} \newline\texttt{\textcolor{grey}{ggformula::}gf\_histogram()} \newline\texttt{\textcolor{grey}{ggformula::}gf\_point()} \newline\texttt{\textcolor{grey}{mosaic::}mean()} \newline\texttt{\textcolor{grey}{mosaic::}median()} \newline\texttt{\textcolor{grey}{base::}options()} \newline\texttt{\textcolor{grey}{mosaic::}pdata()} \newline\texttt{\textcolor{grey}{mosaic::}prop.test()} \newline\texttt{\textcolor{grey}{utils::}read.csv()} \newline\texttt{\textcolor{grey}{mosaic::}resample()} \newline\texttt{\textcolor{grey}{mosaic::}sd()} \newline\texttt{\textcolor{grey}{mosaic::}shuffle()} \newline\texttt{\textcolor{grey}{mosaic::}t.test()} \newline\texttt{\textcolor{grey}{mosaic::}tally()} \newline\texttt{\textcolor{grey}{base::}transform()} \newline\linespread{2}
\vspace{3mm}\setlength{\parindent}{15pt}

\noindent
Used only in formula.
\columnbreak

\noindent
\linespread{1}
\noindent \newline\texttt{\textcolor{grey}{ggplot2::}aes()}\newline\texttt{\textcolor{grey}{forcats::}as\_factor()}\newline\texttt{\textcolor{grey}{base::}c()}\newline\texttt{\textcolor{grey}{infer::}calculate()}\newline\texttt{\textcolor{grey}{infer::}chisq\_test()}\newline\texttt{\textcolor{grey}{stats::}cor()}\newline\texttt{\textcolor{grey}{tidyr::}drop\_na()}\newline\texttt{\textcolor{grey}{stats::}fivenum()}\newline\texttt{\textcolor{grey}{infer::}generate()}\newline\texttt{\textcolor{grey}{ggplot2::}geom\_bar()}\newline\texttt{\textcolor{grey}{ggplot2::}geom\_boxplot()}\newline\texttt{\textcolor{grey}{ggplot2::}geom\_density()}\newline\texttt{\textcolor{grey}{ggplot2::}geom\_histogram()}\newline\texttt{\textcolor{grey}{ggplot2::}geom\_point()}\newline\texttt{\textcolor{grey}{infer::}get\_ci()}\newline\texttt{\textcolor{grey}{infer::}get\_p\_value()}\newline\texttt{\textcolor{grey}{ggplot2::}ggplot()}\newline\texttt{\textcolor{grey}{dplyr::}group\_by()}\newline\texttt{\textcolor{grey}{utils::}help()}\newline\texttt{\textcolor{grey}{infer::}hypothesize()}\newline\texttt{\textcolor{grey}{stats::}IQR()}\newline\texttt{\textcolor{grey}{base::}max()}\newline\texttt{\textcolor{grey}{base::}mean()}\newline\texttt{\textcolor{grey}{base::}median()}\newline\texttt{\textcolor{grey}{base::}min()}\newline\texttt{\textcolor{grey}{dplyr::}mutate()}\newline\texttt{\textcolor{grey}{dplyr::}n()}\newline\texttt{\textcolor{grey}{infer::}prop\_test()}\newline\texttt{\textcolor{grey}{dplyr::}pull()}\newline\texttt{\textcolor{grey}{stats::}quantile()}\newline\texttt{\textcolor{grey}{readr::}read\_csv()}\newline\texttt{\textcolor{grey}{stats::}sd()}\newline\texttt{\textcolor{grey}{infer::}specify()}\newline\texttt{\textcolor{grey}{base::}sum()}\newline\texttt{\textcolor{grey}{dplyr::}summarize()}\newline\texttt{\textcolor{grey}{infer::}t\_test()}\newline\linespread{2}
\vspace{3mm}\setlength{\parindent}{15pt}

\noindent
Used only in \texttt{tidyverse}.
\vfill
 \raggedcolumns
\end{multicols*}
\end{singlespace}

\bibliographystyle{agsm}
\bibliography{references.bib}

\end{document}